\documentclass{article}
\usepackage{arxiv}
\usepackage{hyperref}
\usepackage{latexsym}
\usepackage{url}
\usepackage[utf8]{inputenc}
\usepackage{booktabs}
\usepackage{amsmath}
\usepackage{amsfonts}
\usepackage{amssymb}
\usepackage[inline]{enumitem}
\usepackage{enumitem}
\usepackage{mathbbol}
\usepackage[T1]{fontenc}
\usepackage{mathtools}
\usepackage{multirow}
\usepackage{natbib}
\usepackage{rotating}
\usepackage{subcaption}
\usepackage{graphicx}
\usepackage{etaremune}
\usepackage{authblk}
\usepackage{tikz}

\newcommand{\ie}{\emph{i.e.}}
\newcommand{\eg}{\emph{e.g.}}
\newcommand{\wrt}{\emph{w.r.t. }}

\newcommand{\vs}{\emph{vs. }}

\title{Overview of the TREC 2019 deep learning track}

\author[1]{Nick Craswell}
\author[1]{Bhaskar Mitra}
\author[2]{Emine Yilmaz}
\author[1]{Daniel Campos}
\author[3]{Ellen M. Voorhees}
\affil[1]{Microsoft AI \& Research\\\texttt{\small \{nickcr, bmitra, dacamp\}@microsoft.com}}
\affil[2]{University College London\\\texttt{\small emine.yilmaz@ucl.ac.uk}}
\affil[3]{NIST\\\texttt{\small Ellen.Voorhees@nist.gov}}

\begin{document}
\maketitle

\begin{abstract}

The Deep Learning Track is a new track for TREC 2019, with the goal of studying ad hoc ranking in a large data regime. It is the first track with large human-labeled training sets, introducing two sets corresponding to two tasks, each with rigorous TREC-style blind evaluation and reusable test sets. The document retrieval task has a corpus of 3.2 million documents with 367 thousand training queries, for which we generate a reusable test set of 43 queries. The passage retrieval task has a corpus of 8.8 million passages with 503 thousand training queries, for which we generate a reusable test set of 43 queries. This year 15 groups submitted a total of 75 runs, using various combinations of deep learning, transfer learning and traditional IR ranking methods. Deep learning runs significantly outperformed traditional IR runs. Possible explanations for this result are that we introduced large training data and we included deep models trained on such data in our judging pools, whereas some past studies did not have such training data or pooling.

\end{abstract}
\section{Introduction}
\label{sec:intro}

Deep learning methods, where a computational model learns an intricate representation of a large-scale dataset, have yielded dramatic improvements on the state of the art in speech recognition and computer vision. This has been fueled by the availability of large-scale datasets \citep{lecun2015deep} such as the ImageNet dataset \citep{deng2009imagenet} for computer vision and the Atari Arcade Learning Environment \citep{bellemare2013arcade} for game playing.

There has been significant interest in deep learning for ad-hoc ranking~\citep{mitra2018introduction}. Work so far has largely been done with small data, proprietary data or synthetic data. With small data, there has been some discussion about whether deep learning methods really outperform strong traditional IR baselines \citep{Yang2019critically}. Using a proprietary set of document ranking data with 200,000 training queries~\citep{mitra2017learning}, a traditional IR baseline was beaten, but it was impossible for others to follow up on the work without a data release. \cite{dietz2017trec} have a TREC task with enough training data to investigate such findings, but on synthetic rather than human-labeled data. 

Since significant questions remain about baselines and the required volume of human-labeled data, we argue that TREC is a good forum for studying such issues. When a large human-labeled dataset is made available, participants can investigate the role of data size by subsampling. Strong baselines are more than welcome at TREC and there is a blind one-shot evaluation to avoid overfitting.

The TREC 2019 Deep Learning Track has two tasks: Document retrieval and passage retrieval. Each task has a dataset that is new to TREC, although the passage task is similar to the MS MARCO passage ranking leaderboard~\citep{bajaj2016ms}, but with a new test set in the TREC version with more comprehensive labeling. Both tasks are ad-hoc retrieval, meaning that there is a fixed document set, and the goal of the information retrieval system is to respond to each new query with results that would satisfy the querying user's information need. Ad-hoc retrieval is a very common scenario in real-world search applications and in TREC.

The main goals of the track are:
\begin{enumerate*}[label=\arabic*)]
\item To provide large reusable datasets for training and evaluation of deep learning and traditional ranking methods in a large training data regime, 
\item To perform a rigorous blind single-shot evaluation, where test labels don't even exist until after all runs are submitted, to compare different ranking methods, and
\item To study this in both a traditional TREC setup with end-to-end retrieval and in a re-ranking setup that matches how some models may be deployed in practice.
\end{enumerate*}

\paragraph{Comparing ad hoc retrieval methods in a large-data regime.}
The track should help us build our understanding of how retrieval methods can take advantage of large-scale data. It should also allow participants to compare various ranking methods such as:
\begin{itemize}
\item ML models \vs traditional IR---including pseudo-relevance feedback.
\item Deep learning \vs feature-based learning-to-rank (LTR) methods \citep{Liu:2009}. 
\item Comparison of different deep learning architectures. 
\item Comparison of different supervision approaches, such as fully supervised \vs semi-supervised \vs weakly supervised deep learning \citep{dehghani2017neural}.
\item Comparison of such models with all the training labels \vs using a subset of labels, to see how performance improves with more data.
\end{itemize}
Comparing different methods for ad hoc search has always been a focus area at TREC, so our goal in this track is to continue that work.


\paragraph{End-to-end retrieval \vs reranking.}
In real-world implementations of LTR methods, a common technique is to first retrieve the top-$k$ documents for a query using relatively cheap ``phase 1'' ranker such as BM25, and then apply the full ML model to rerank the top-$k$ documents in ``phase 2''.

This motivates us to offer two participation styles in the Deep Learning Track, which we also refer to as subtasks.
One is to implement full end-to-end retrieval, perhaps by implementing both phase 1 and phase 2.
This is interesting because a good implementation of phase 1 can enhance the end-to-end performance of the system, by enriching the candidate set for phase 2.
It also encourages participants to consider alternatives to the two-phase approach, if it can improve efficiency and effectiveness.

The other participation style is to only implement a top-$k$ reranker.
This approach is realistic in practice, in fact it is simply phase 2 of the end-to-end approach, for a fixed phase 1.
This style of participation lowers the barrier to entry for participating groups who are interested in the LTR aspects of dealing with a large number of training queries, but are not interested in indexing a corpus or studying phase 1 issues.
In this style of evaluation---sometimes referred to as \emph{telescoping} \citep{matveeva2006high}---participants are given the top-$k$ results in both the training and test set.

The interaction between deep learning models and traditional IR indexing data structures is also particularly interesting.
Most applications of deep learning models in IR---with few exceptions \eg, \citep{boytsov2016off, zamani2018neural, mitra2019incorporating, nogueira2019document}---have been constrained to the reranking setting.
Encouraging future exploration of deep learning based ranking models under the full retrieval settings is an explicit goal of the Deep Learning Track.

\section{Task description}
\label{sec:task}

The track has two tasks: Document retrieval and passage retrieval.
Participants were allowed to submit up to three runs per task, although this was not strictly enforced.
Participants were provided with an initial set of $200$ test queries, then NIST later selected $43$ queries during the pooling and judging process, based on budget constraints and with the goal of producing a reusable test collection.
The same $200$ queries were used for submissions in both tasks, while the selected $43$ queries for each task were overlapping but not identical. The full judging process is described in Section~\ref{sec:reusability}.

When submitting each run, participants also indicated what external data, pretrained models and other resources were used, as well as information on what style of model was used.
Below we provide more detailed information about the document retrieval and passage retrieval tasks, as well as the datasets provided as part of these tasks. 

\subsection{Document retrieval task}

The first task focuses on document retrieval---with two subtasks:
\begin{enumerate*}[label=(\roman*)]
    \item Full retrieval and
    \item top-$100$ reranking.
\end{enumerate*}

In the full retrieval subtask, the runs are expected to rank documents based on their relevance to the query, where documents can be retrieved from the full document collection provided.
This subtask models the end-to-end retrieval scenario.
Note, although most full retrieval runs had 1000 results per query, the reranking runs had 100, so to make the AP and RR results more comparable across subtasks we truncated full retrieval runs by taking the top-100 results per query by score. These truncated runs were used in the main results table for the task (only), not in the TREC Appendix or in Section~\ref{sec:reusability}. 

In the reranking subtask, participants were provided with an initial ranking of $100$ documents, giving all participants the same starting point. The 100 were retrieved using Indri \citep{strohman2005indri} on the full corpus with Krovetz stemming and stopwords eliminated. Participants were expected to rerank the candidates \wrt their estimated relevance to the query.
This is a common scenario in many real-world retrieval systems that employ a telescoping architecture \citep{matveeva2006high, wang2011cascade}.
The reranking subtask allows participants to focus on learning an effective relevance estimator, without the need for implementing an end-to-end retrieval system.
It also makes the reranking runs more comparable, because they all rerank the same set of 100 candidates.

For judging, NIST's pooling was across both subtasks, and they also identified additional documents for judging via classifier. Further, for queries with many relevant documents, additional documents were judged. These steps were carried out to identify a sufficiently comprehensive set of relevant results, to allow reliable future dataset reuse. Judgments were on a four-point scale:
\begin{etaremune}[start=3]
    \item \textbf{Perfectly relevant:} Document is dedicated to the query, it is worthy of being a top result in a search engine.
    \item \textbf{Highly relevant:} The content of this document provides substantial information on the query.
    \item \textbf{Relevant:} Document provides some information relevant to the query, which may be minimal.
    \item \textbf{Irrelevant:} Document does not provide any useful information about the query.
\end{etaremune}

\subsection{Passage retrieval task}

Similar to the document retrieval task, the passage retrieval task includes
\begin{enumerate*}[label=(\roman*)]
    \item a full retrieval and
    \item a top-$1000$ reranking tasks.
\end{enumerate*}

In the full retrieval subtask, given a query, the participants were expected to retrieve a ranked list of passages from the full collection based on their estimated likelihood of containing an answer to the question.
Participants could submit up to 1000 passages per query for this end-to-end retrieval task.

In the top-$1000$ reranking subtask, 1000 passages per query query were provided to participants, giving all participants the same starting point. The sets of 1000 were generated based on BM25 retrieval with no stemming as applied to the full collection. Participants were expected to rerank the 1000 passages based on their estimated likelihood of containing an answer to the query.
In this subtask, we can compare different reranking methods based on the same initial set of $1000$ candidates, with the same rationale as described for the document reranking subtask.

For judging, NIST's pooling was across both subtasks, and they also identified additional passages for judging via classifier. Further, for queries with many relevant passages, additional passages were judged. These steps were carried out to identify a sufficiently comprehensive set of relevant results, to allow reliable future dataset reuse. Judgments were on a four-point scale:
\begin{etaremune}[start=3]
    \item \textbf{Perfectly relevant:} The passage is dedicated to the query and contains the exact answer.
    \item \textbf{Highly relevant:} The passage has some answer for the query, but the answer may be a bit unclear, or hidden amongst extraneous information.
    \item \textbf{Related:} The passage seems related to the query but does not answer it.
    \item \textbf{Irrelevant:} The passage has nothing to do with the query.
\end{etaremune}


\section{Datasets}
\label{sec:data}

Both tasks have large training sets based on human relevance assessments, derived from MS MARCO.
These are sparse, with no negative labels and often only one positive label per query, analogous to some real-world training data such as click logs.

In the case of passage retrieval, the positive label indicates that the passage contains an answer to a query. 
In the case of document retrieval, we transferred the passage-level label to the corresponding source document that contained the passage. We do this under the assumption that a document with a relevant passage is a relevant document, although we note that our document snapshot was generated at a different time from the passage dataset, so there can be some mismatch. Despite this, in this year's document retrieval task machine learning models seem to benefit from using the labels, when evaluated using NIST's non-sparse, non-transferred labels. This suggests the transferred document labels are meaningful for our TREC task.

The passage corpus is the same as in MS MARCO passage retrieval leaderboard. 
The document corpus is newly released for use in TREC. 
Each document has three fields:
\begin{enumerate*}[label=(\roman*)]
    \item URL,
    \item title, and
    \item body text.
\end{enumerate*}

\begin{table}
    \centering
    \caption{Summary of statistics on TREC 2019 Deep Learning Track datasets.}
    \begin{tabular}{lrrrrrr}
    \hline
    \hline
        & & \multicolumn{2}{c}{\textbf{Document retrieval dataset}} & & \multicolumn{2}{c}{\textbf{Passage retrieval dataset}} \\
        \textbf{File description} & & Number of records & File size & & Number of records & File size \\
        \hline
        Collection & & $3,213,835$ & $22$ GB & & $8,841,823	$ & $2.9$ GB \\
        Train queries & & $367,013$ & $15$ MB & & $502,940$ & $19.7$ MB \\
        Train qrels & & $384,597$ & $7.6$ MB & & $532,761$ & $10.1$ MB \\
        Validation queries & & $5,193$ & $216$ KB & & $12,665$ & $545$ KB \\
        Validation qrels & & $519,300$ & $27$ MB & & $59,273$ & $1.1$ MB \\
        Test queries & & $200$ & $12$ KB & & $200$ & $12$ KB \\
        \hline
        \hline
    \end{tabular}
    \label{tbl:data}
\end{table}

Table \ref{tbl:data} provides descriptive statistics for the datasets.
More details about the datasets---including directions for download---is available on the TREC 2019 Deep Learning Track website\footnote{\url{https://microsoft.github.io/TREC-2019-Deep-Learning/}}.
Interested readers are also encouraged to refer to \citep{bajaj2016ms} for details on the original MS MARCO dataset.
\section{Results and analysis}
\label{sec:result}

\begin{table}
    \centering
    \caption{Summary of statistics of runs for the two retrieval tasks at the TREC 2019 Deep Learning Track.}
    \begin{tabular}{lll}
    \hline
    \hline
        & \textbf{Document retrieval} & \textbf{Passage retrieval} \\
        \hline
        Number of groups & 10 & 11 \\
        Number of total runs & 38 & 37 \\
        Number of runs w/ category: nnlm & 15 & 18 \\
        Number of runs w/ category: nn & 12 & 8 \\
        Number of runs w/ category: trad & 11 & 11 \\
        Number of runs w/ category: rerank & 10 & 11 \\
        Number of runs w/ category: fullrank & 28 & 26 \\
        \hline
        \hline
    \end{tabular}
    \label{tbl:runs-by-type}
\end{table}

\paragraph{Submitted runs}
A total of 15 groups participated in the TREC 2019 Deep Learning Track, with an aggregate of 75 runs submitted across both tasks.

Based run submission surveys, we classify each run into one of three categories:
\begin{itemize}
    \item \textbf{nnlm:} if the run employs large scale pre-trained neural language models, such as BERT \citep{devlin2018bert} or XLNet \citep{yang2019xlnet}
    \item \textbf{nn:} if the run employs some form of neural network based approach---\eg, Duet \citep{mitra2017learning, mitra2019updated} or using word embeddings \citep{joulin2016bag}---but does not fall into the ``nnlm'' category
    \item \textbf{trad:} if the run exclusively uses traditional IR methods like BM25 \citep{robertson2009probabilistic} and RM3 \citep{abdul2004umass}.
\end{itemize}
We placed 33 ($44\%$) runs in the ``nnlm'' category (32 using BERT and one using XLNet), 20 ($27\%$) in the ``nn'' category, and the remaining 22 ($29\%$) in the ``trad'' category.

We further categorize runs based on subtask:
\begin{itemize}
    \item \textbf{rerank:} if the run reranks the provided top-$k$ candidates, or
    \item \textbf{fullrank:} if the run employs their own phase 1 retrieval system.
\end{itemize}
We find that only 21 ($28\%$) submissions fall under the ``rerank'' category---while the remaining 54 ($72\%$) are ``fullrank''.
Table~\ref{tbl:runs-by-type} breaks down the submissions by category and task.

We also encouraged some participants to run strong traditional IR baselines, and submit them as additional runs under the ``BASELINE'' group. Baseline runs for document ranking were:
\begin{description}[align=right,labelwidth=3cm]
\item[bm25base] BM25 \citep{robertson2009probabilistic} with default parameters
\item[bm25base\_ax] BM25+AX \citep{yang2019reproducing} with default parameters
\item[bm25base\_prf] BM25+PRF \citep{sakai2019prf} with default parameters
\item[bm25base\_rm3] BM25+RM3 \citep{Yang2019critically} with default parameters
\item[bm25tuned] BM25 \citep{robertson2009probabilistic} with tuned parameters
\item[bm25tuned\_ax] BM25+AX \citep{yang2019reproducing} with tuned parameters
\item[bm25tuned\_prf] BM25+PRF \citep{sakai2019prf} with tuned parameters
\item[bm25tuned\_rm3] BM25+RM3 \citep{Yang2019critically} with tuned parameters
\end{description}
Baseline runs for passage ranking were:
\begin{description}[align=right,labelwidth=3cm]
\item[bm25base\_ax\_p] BM25+AX \citep{yang2019reproducing} with default parameters
\item[bm25base\_p] BM25 \citep{robertson2009probabilistic} with default parameters
\item[bm25base\_prf\_p] BM25+PRF \citep{sakai2019prf} with default parameters
\item[bm25base\_rm3\_p] BM25+RM3 \citep{Yang2019critically} with default parameters
\item[bm25tuned\_ax\_p] BM25+AX \citep{yang2019reproducing} with tuned parameters
\item[bm25tuned\_p] BM25 \citep{robertson2009probabilistic} with tuned parameters
\item[bm25tuned\_prf\_p] BM25+PRF \citep{sakai2019prf} with tuned parameters
\item[bm25tuned\_rm3\_p] BM25+RM3 \citep{Yang2019critically} with tuned parameters
\end{description}

\begin{table}
    \small
    \centering
    \caption{Document retrieval runs. RR (MS) is based on MS MARCO labels. All other metrics are based on NIST labels.}
\begin{tabular}{llllrrrrr}
\toprule
run &         group &   subtask & neural &  RR (MS) &     RR &  NDCG@10 & NCG@100 & AP \\
\midrule
idst\_bert\_v3    &          IDST &  fullrank &   nnlm &   0.4866 & 0.9612 &   0.7257 & 0.5800 & 0.3137 \\
idst\_bert\_r1    &          IDST &    rerank &   nnlm &   0.4889 & 0.9729 &   0.7189 & 0.5179 & 0.2915 \\
idst\_bert\_v2    &          IDST &  fullrank &   nnlm &   0.4865 & 0.9612 &   0.7181 & 0.5947	 & 0.3157 \\
idst\_bert\_v1    &          IDST &  fullrank &   nnlm &   0.4874 & 0.9729 &   0.7175 & 0.5820 & 0.3119 \\
idst\_bert\_r2    &          IDST &    rerank &   nnlm &   0.4734 & 0.9729 &   0.7135 & 0.5179 & 0.2910 \\
bm25exp\_marcomb &        h2oloo &  fullrank &   nnlm &   0.3518 & 0.8992 &   0.6456 & 0.6367 & 0.3190 \\
TUW19-d3-re     &     TU-Vienna &    rerank &     nn &   0.4014 & 0.9457 &   0.6443 & 0.5179 & 0.2709 \\
ucas\_runid1     &          UCAS &    rerank &   nnlm &   0.4422 & 0.9109 &   0.6437 & 0.5179 & 0.2642 \\
ucas\_runid3     &          UCAS &    rerank &   nnlm &   0.4353 & 0.8992 &   0.6418 & 0.5179 & 0.2677 \\
bm25\_marcomb    &        h2oloo &  fullrank &   nnlm &   0.3591 & 0.9128 &   0.6403 & 0.6356 & 0.3229 \\
bm25exp\_marco   &        h2oloo &  fullrank &   nnlm &   0.3610 & 0.9031 &   0.6399 & 0.6191 & 0.3030 \\
ucas\_runid2     &          UCAS &    rerank &   nnlm &   0.4315 & 0.9496 &   0.6350 & 0.5179 & 0.2526 \\
TUW19-d2-re     &     TU-Vienna &    rerank &     nn &   0.3154 & 0.9147 &   0.6053 & 0.5179 & 0.2391 \\
uogTrDNN6LM     &         uogTr &  fullrank &   nnlm &   0.3187 & 0.8729 &   0.6046 & 0.5093 & 0.2488 \\
TUW19-d1-re     &     TU-Vienna &    rerank &     nn &   0.3616 & 0.8915 &   0.5930 & 0.5179 & 0.2524 \\
ms\_ensemble     &     Microsoft &  fullrank &     nn &   0.3725 & 0.8760 &   0.5784 & 0.4841 & 0.2369 \\
srchvrs\_run1    &       srchvrs &  fullrank &   trad &   0.3065 & 0.8715 &   0.5609 & 0.5599 & 0.2645 \\
TUW19-d2-f      &     TU-Vienna &  fullrank &     nn &   0.2886 & 0.8711 &   0.5596 & 0.4103 & 0.2050 \\
TUW19-d3-f      &     TU-Vienna &  fullrank &     nn &   0.3735 & 0.8929 &   0.5576 & 0.3045 & 0.1843 \\
dct\_tp\_bm25e2   &           CMU &  fullrank &     nn &   0.3402 & 0.8718 &   0.5544 & 0.4979 & 0.2244 \\
srchvrs\_run2    &       srchvrs &  fullrank &   trad &   0.3038 & 0.8715 &   0.5529 & 0.5572 & 0.2615 \\
bm25tuned\_rm3   &      BASELINE &  fullrank &   trad &   0.3396 & 0.8074 &   0.5485 & 0.5590 & 0.2700 \\
dct\_qp\_bm25e    &           CMU &  fullrank &     nn &   0.3585 & 0.8915 &   0.5435 & 0.4924 & 0.2228 \\
dct\_tp\_bm25e    &           CMU &  fullrank &     nn &   0.3530 & 0.8638 &   0.5424 & 0.4786 & 0.2098 \\
uogTrDSSQE5LM   &         uogTr &  fullrank &   nnlm &   0.3264 & 0.8895 &   0.5386 & 0.1839 & 0.1085 \\
TUW19-d1-f      &     TU-Vienna &  fullrank &     nn &   0.3190 & 0.8465 &   0.5383 & 0.2951 & 0.1647 \\
ms\_duet         &     Microsoft &    rerank &     nn &   0.2758 & 0.8101 &   0.5330 & 0.5179 & 0.2291 \\
uogTrDSS6pLM    &         uogTr &  fullrank &   nnlm &   0.2803 & 0.8895 &   0.5323 & 0.1868 & 0.1129 \\
bm25tuned\_prf   &      BASELINE &  fullrank &   trad &   0.3176 & 0.8005 &   0.5281 & 0.5576 & 0.2759 \\
bm25tuned\_ax    &      BASELINE &  fullrank &   trad &   0.2889 & 0.7492 &   0.5245 & 0.5835 & 0.2816 \\
bm25base        &      BASELINE &  fullrank &   trad &   0.2949 & 0.8046 &   0.5190 & 0.5170 & 0.2443 \\
bm25base\_rm3    &      BASELINE &  fullrank &   trad &   0.2405 & 0.7714 &   0.5169 & 0.5546 & 0.2772 \\
runid1          &  CCNU\_IRGroup &    rerank &   nnlm &   0.3058 & 0.7811 &   0.5164 & 0.5179 & 0.2366 \\
bm25tuned       &      BASELINE &  fullrank &   trad &   0.2930 & 0.8872 &   0.5140 & 0.5262 & 0.2318 \\
bm25base\_prf    &      BASELINE &  fullrank &   trad &   0.2717 & 0.7774 &   0.5106 & 0.5303 & 0.2542 \\
baseline        &      BITEM\_DL &  fullrank &   trad &   0.2795 & 0.8037 &   0.4823 & 0.5114 & 0.2168 \\
bm25base\_ax     &      BASELINE &  fullrank &   trad &   0.2677 & 0.7424 &   0.4730 & 0.5148 & 0.2452 \\
\bottomrule
\end{tabular}

    \label{tbl:results-docs}
\end{table}

\begin{table}
    \small
    \centering
    \caption{Passage retrieval runs. RR (MS) is based on MS MARCO labels. All other metrics are based on NIST labels.}
\begin{tabular}{llllrrrrr}
\toprule
run &         group &   subtask & neural &  RR (MS) &     RR &  NDCG@10 & NCG@1000 &   AP \\
\midrule
idst\_bert\_p1    &          IDST &  fullrank &   nnlm &   0.4635 & 0.9283 &   0.7645 & 0.8196 & 0.5030 \\
idst\_bert\_p2    &          IDST &  fullrank &   nnlm &   0.4631 & 0.9283 &   0.7632 & 0.8203 & 0.5039 \\
idst\_bert\_p3    &          IDST &  fullrank &   nnlm &   0.4374 & 0.9167 &   0.7594 & 0.8287 & 0.5046 \\
p\_exp\_rm3\_bert  &        h2oloo &  fullrank &   nnlm &   0.3582 & 0.8884 &   0.7422 & 0.7939 & 0.5049 \\
p\_bert          &        h2oloo &  fullrank &   nnlm &   0.3624 & 0.8663 &   0.7380 & 0.7472 & 0.4677 \\
idst\_bert\_pr2   &          IDST &    rerank &   nnlm &   0.4209 & 0.8818 &   0.7379 & 0.6864 & 0.4565 \\
idst\_bert\_pr1   &          IDST &    rerank &   nnlm &   0.4430 & 0.9070 &   0.7378 & 0.6864 & 0.4571 \\
p\_exp\_bert      &        h2oloo &  fullrank &   nnlm &   0.3564 & 0.8671 &   0.7336 & 0.7465 & 0.4749 \\
test1           &         Brown &    rerank &   nnlm &   0.3598 & 0.8702 &   0.7314 & 0.6864 & 0.4567 \\
TUA1-1          &          TUA1 &    rerank &   nnlm &   0.3622 & 0.8702 &   0.7314 & 0.6864 & 0.4571 \\
runid4          &     udel\_fang &    rerank &   nnlm &   0.3762 & 0.8702 &   0.7028 & 0.6864 & 0.4383 \\
runid3          &     udel\_fang &    rerank &   nnlm &   0.3725 & 0.8663 &   0.6975 & 0.6864 & 0.4381 \\
TUW19-p3-f      &     TU-Vienna &  fullrank &     nn &   0.3134 & 0.8407 &   0.6884 & 0.7436 & 0.4196 \\
TUW19-p1-f      &     TU-Vienna &  fullrank &     nn &   0.3187 & 0.8360 &   0.6756 & 0.7436 & 0.4125 \\
TUW19-p3-re     &     TU-Vienna &    rerank &     nn &   0.3100 & 0.8568 &   0.6746 & 0.6864 & 0.4113 \\
TUW19-p1-re     &     TU-Vienna &    rerank &     nn &   0.3180 & 0.8516 &   0.6746 & 0.6864 & 0.4073 \\
TUW19-p2-f      &     TU-Vienna &  fullrank &     nn &   0.3469 & 0.8487 &   0.6709 & 0.7432 & 0.4157 \\
ICT-BERT2       &        ICTNET &  fullrank &   nnlm &   0.3846 & 0.8743 &   0.6650 & 0.2491 & 0.2421 \\
srchvrs\_ps\_run2 &       srchvrs &  fullrank &   nnlm &   0.3262 & 0.8302 &   0.6645 & 0.6643 & 0.4090 \\
TUW19-p2-re     &     TU-Vienna &    rerank &     nn &   0.3424 & 0.8611 &   0.6615 & 0.6864 & 0.3963 \\
ICT-CKNRM\_B     &        ICTNET &  fullrank &   nnlm &   0.2984 & 0.8016 &   0.6481 & 0.2491 & 0.2289 \\
ms\_duet\_passage &     Microsoft &    rerank &     nn &   0.2473 & 0.8065 &   0.6137 & 0.6864 & 0.3477 \\
ICT-CKNRM\_B50   &        ICTNET &  fullrank &   nnlm &   0.2055 & 0.7597 &   0.6014 & 0.3786 & 0.2429 \\
srchvrs\_ps\_run3 &       srchvrs &  fullrank &   trad &   0.1883 & 0.6942 &   0.5558 & 0.7240 & 0.3184 \\
bm25tuned\_prf\_p &      BASELINE &  fullrank &   trad &   0.1928 & 0.6996 &   0.5536 & 0.7947 & 0.3684 \\
bm25base\_ax\_p   &      BASELINE &  fullrank &   trad &   0.1888 & 0.6516 &   0.5511 & 0.8194 & 0.3745 \\
bm25tuned\_ax\_p  &      BASELINE &  fullrank &   trad &   0.1840 & 0.6481 &   0.5461 & 0.8145 & 0.3632 \\
bm25base\_prf\_p  &      BASELINE &  fullrank &   trad &   0.2007 & 0.6211 &   0.5372 & 0.7901 & 0.3561 \\
runid2          &  CCNU\_IRGroup &    rerank &   nnlm &   0.2143 & 0.8088 &   0.5322 & 0.6830 & 0.2671 \\
runid5          &  CCNU\_IRGroup &  fullrank &   nnlm &   0.2068 & 0.7999 &   0.5252 & 0.5440 & 0.2506 \\
bm25tuned\_rm3\_p &      BASELINE &  fullrank &   trad &   0.2162 & 0.6992 &   0.5231 & 0.7841 & 0.3377 \\
bm25base\_rm3\_p  &      BASELINE &  fullrank &   trad &   0.1590 & 0.6683 &   0.5180 & 0.7976 & 0.3390 \\
bm25base\_p      &      BASELINE &  fullrank &   trad &   0.2402 & 0.7036 &   0.5058 & 0.7490 & 0.3013 \\
srchvrs\_ps\_run1 &       srchvrs &  fullrank &   trad &   0.1902 & 0.5597 &   0.4990 & 0.7240 & 0.2972 \\
bm25tuned\_p     &      BASELINE &  fullrank &   trad &   0.2363 & 0.6850 &   0.4973 & 0.7472 & 0.2903 \\
UNH\_bm25        &     TREMA-UNH &  fullrank &   trad &   0.1803 & 0.6036 &   0.4495 & 0.6957 & 0.2566 \\
\bottomrule
\end{tabular}

    \label{tbl:results-passages}
\end{table}

\paragraph{Overall results}

Our main metric in both tasks is Normalized Discounted Cumulative Gain (NDCG)---specifically, NDCG@10, since it makes use of our 4-level judgments and focuses on the first results that users will see. To analyse if any of the fullrank runs recall more relevant candidates in phase 1 compared to those provided for the reranking subtask, we also report Normalized Cumulative Gain (NCG) \citep{rosset2018optimizing} at rank 100 and 1000 for the document and passage ranking tasks, respectively.
We choose to report NCG because it discriminates between recalling documents with different positive relevance grades and is a natural complement to NDCG, our main metric.
Although NCG is not officially supported by trec\_eval, we confirm that it correlates strongly with the recall metric for these analysed runs.
The overall results are presented in Table~\ref{tbl:results-docs} for document retrieval and Table~\ref{tbl:results-passages} for passage retrieval. These tables include multiple metrics and run categories, which we now use in our analysis.

\paragraph{Evaluation of deep learning and traditional ranking methods in a large training data regime}


An important goal of this track is to compare the performance of different types of model, using large human-labeled training sets, for the core IR task of ad-hoc search.  
Indeed this is the first time a TREC-style blind evaluation has been carried out to compare state-of-the-art neural and traditional IR methods.

\begin{figure}
  \center
  \begin{subfigure}{.49\textwidth}
    \includegraphics[width=\textwidth]{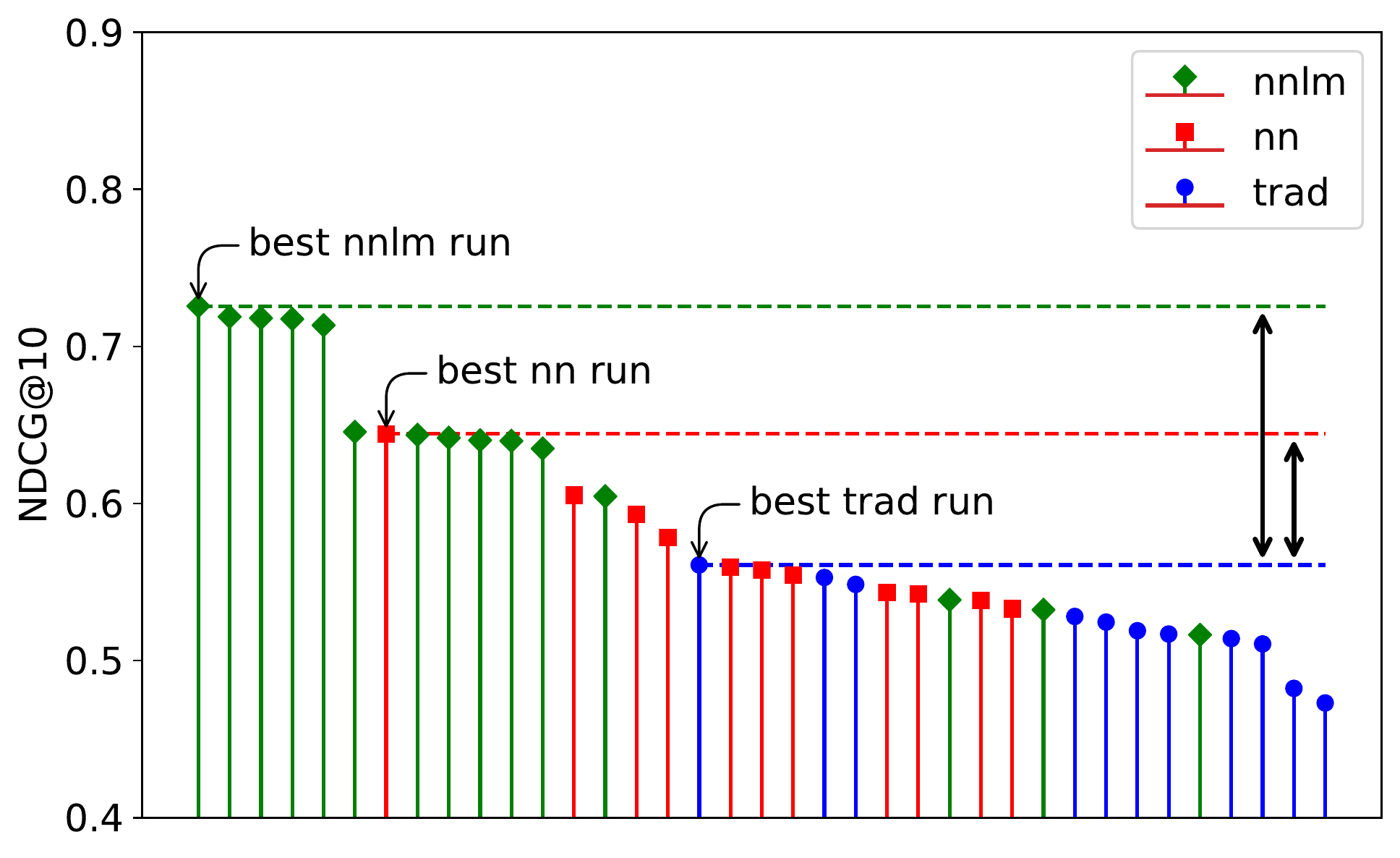}
    \caption{Document retrieval task}
    \label{fig:model-task-docs-stem-by-model-type}
  \end{subfigure}
  \hfill
  \begin{subfigure}{.49\textwidth}
    \includegraphics[width=\textwidth]{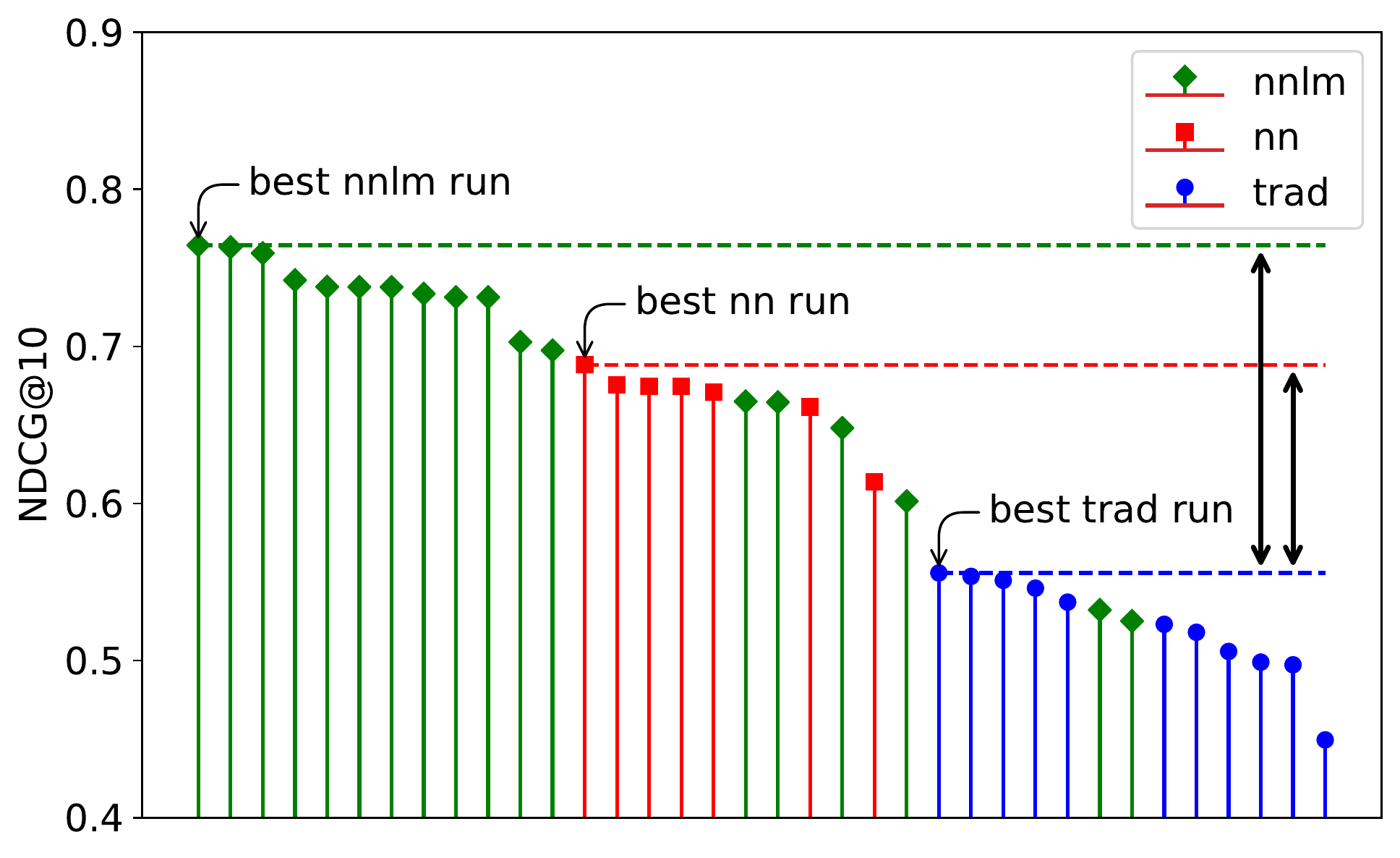}
    \caption{Passage retrieval task}
    \label{fig:model-task-passages-stem-by-model-type}
  \end{subfigure}
  \caption{NDCG@10 results, broken down by run type. Runs of type ``nnlm'', meaning they use language models such as BERT, performed best on both tasks. Other neural network models ``nn'' and non-neural models ``trad'' had relatively lower performance this year. More iterations of evaluation and analysis would be needed to determine if this is a general result, but it is a strong start for the argument that deep learning methods may take over from traditional methods in IR applications.}
  \label{fig:model-stem-by-model-type}
\end{figure}

Figure~\ref{fig:model-task-docs-stem-by-model-type} plots the NDCG@10 performance of the different runs for the document retrieval task, broken down by model type.
In general, runs in the category ``nnlm'' outperform the ``nn'' runs, which outperform the ``trad'' runs.
The best performing run of each category is indicated, with the best ``nnlm'' and ``nn'' models outperforming the best ``trad'' model by $29.4\%$ and $14.8\%$ respectively.

The passage retrieval task reveals similar pattern.
In Figure~\ref{fig:model-task-passages-stem-by-model-type}, the gap between the best ``nnlm'' and ``nn'' runs and the best ``trad'' run is larger, at $37.4\%$ and $23.7\%$ respectively. One explanation for this could be that vocabulary mismatch between queries and relevant results is more likely in short text, so neural methods that can overcome such mismatch have a relatively greater advantage in passage retrieval. Another explanation could be that there is already a public leaderboard, albeit without test labels from NIST, for the passage task. Some TREC participants may have submitted neural models multiple times to the public leaderboard, and are well practiced for the passage ranking task.

\begin{figure}
\includegraphics[width=\textwidth]{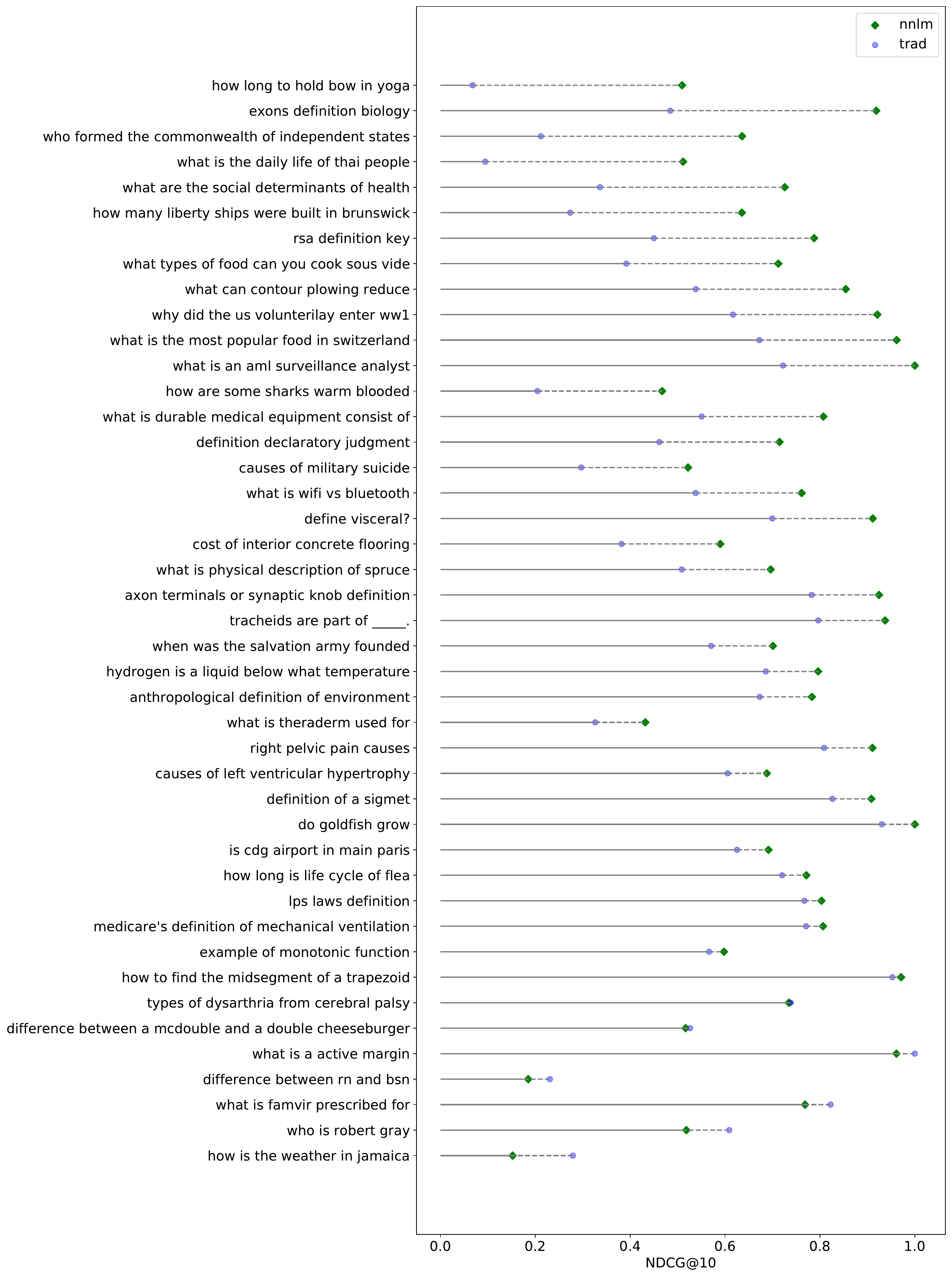}
\caption{Comparison of the best ``nnlm'' and ``trad'' runs on individual test queries for the document retrieval task. Queries are sorted by difference in mean performance between ``nnlm'' and ``trad''runs. Queries on which ``nnlm'' wins with large margin are at the top.}
\label{fig:model-task-docs-bar-per-query}
\end{figure}

\begin{figure}
\includegraphics[width=\textwidth]{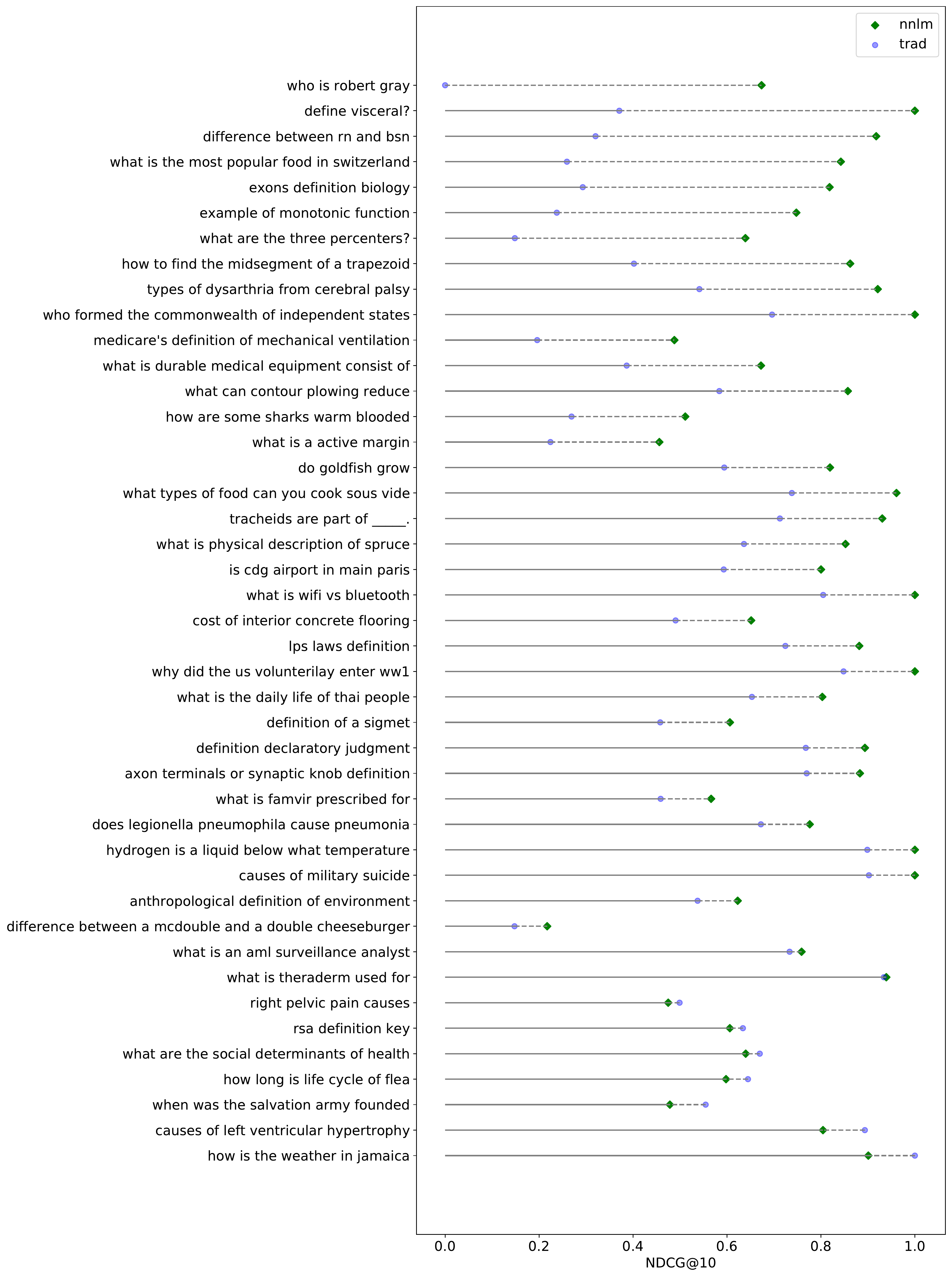}
\caption{Comparison of the best ``nnlm'' and ``trad'' runs on individual test queries for the passage retrieval task. Queries are sorted by difference in mean performance between ``nnlm'' and ``trad''runs. Queries on which ``nnlm'' wins with large margin are at the top.}
\label{fig:model-task-passages-bar-per-query}
\end{figure}

In query-level win-loss analysis for the document retrieval task (Figure~\ref{fig:model-task-docs-bar-per-query}) the best ``nnlm'' model outperforms the best ``trad'' run on 36 out of 43 test queries (\ie, $83.7\%$). Passage retrieval shows a similar pattern in Figure~\ref{fig:model-task-passages-bar-per-query}. Neither task has a large class of queries where the ``nnlm'' model performs worse, at least on this year's data. However, more iterations of rigorous blind evaluation with strong ``trad'' baselines, plus more scrutiny of the benchmarking methods, would be required to convince us that this is true in general.


\begin{figure}
  \center
  \begin{subfigure}{.4\textwidth}
    \includegraphics[width=\textwidth]{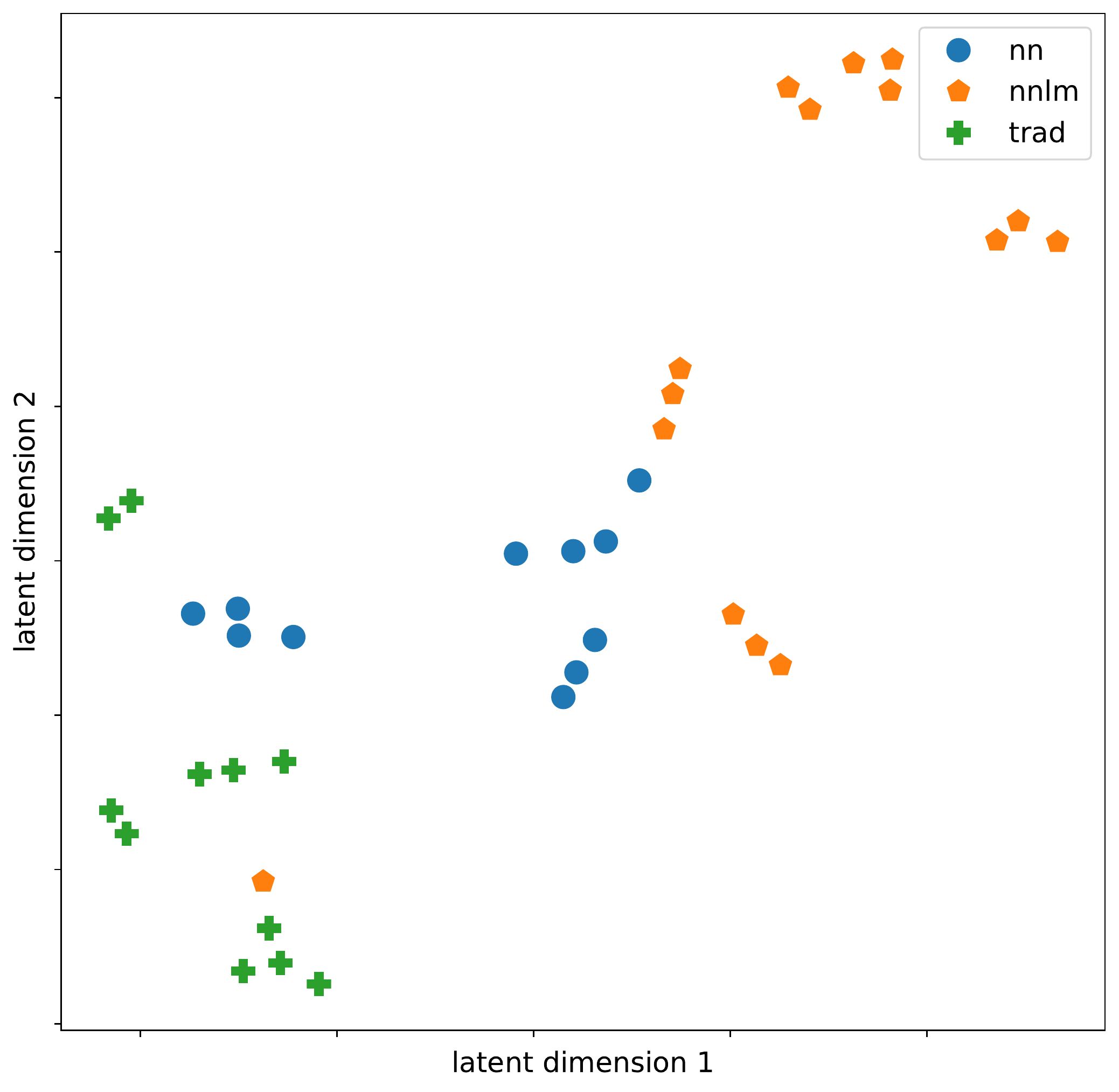}
    \caption{By model type on document retrieval task}
    \label{fig:model-task-docs-by-model-type}
  \end{subfigure}
  \begin{subfigure}{.4\textwidth}
    \includegraphics[width=\textwidth]{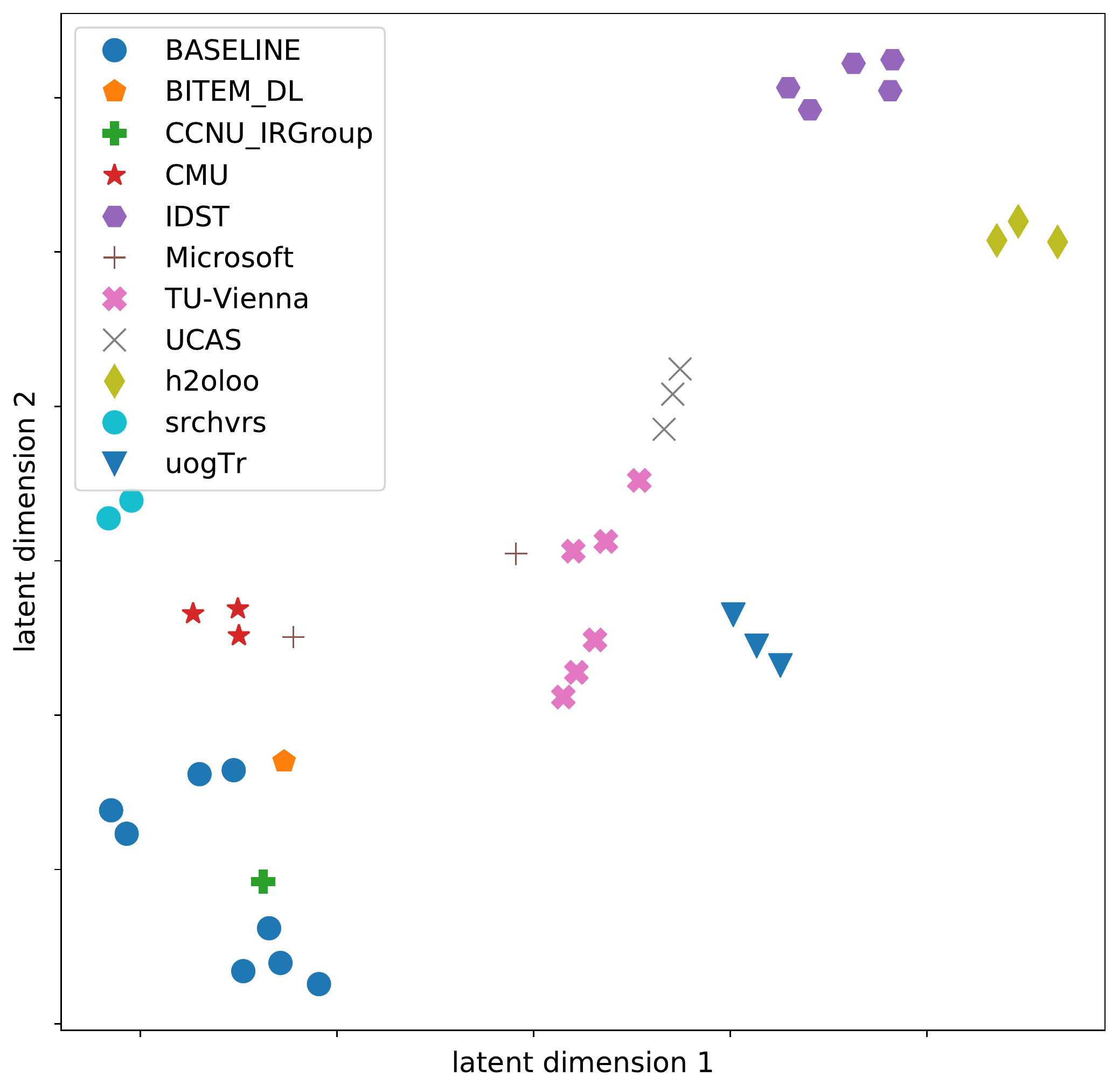}
    \caption{By group name on document retrieval task}
    \label{fig:model-task-docs-by-team-name}
  \end{subfigure}
  \begin{subfigure}{.4\textwidth}
    \includegraphics[width=\textwidth]{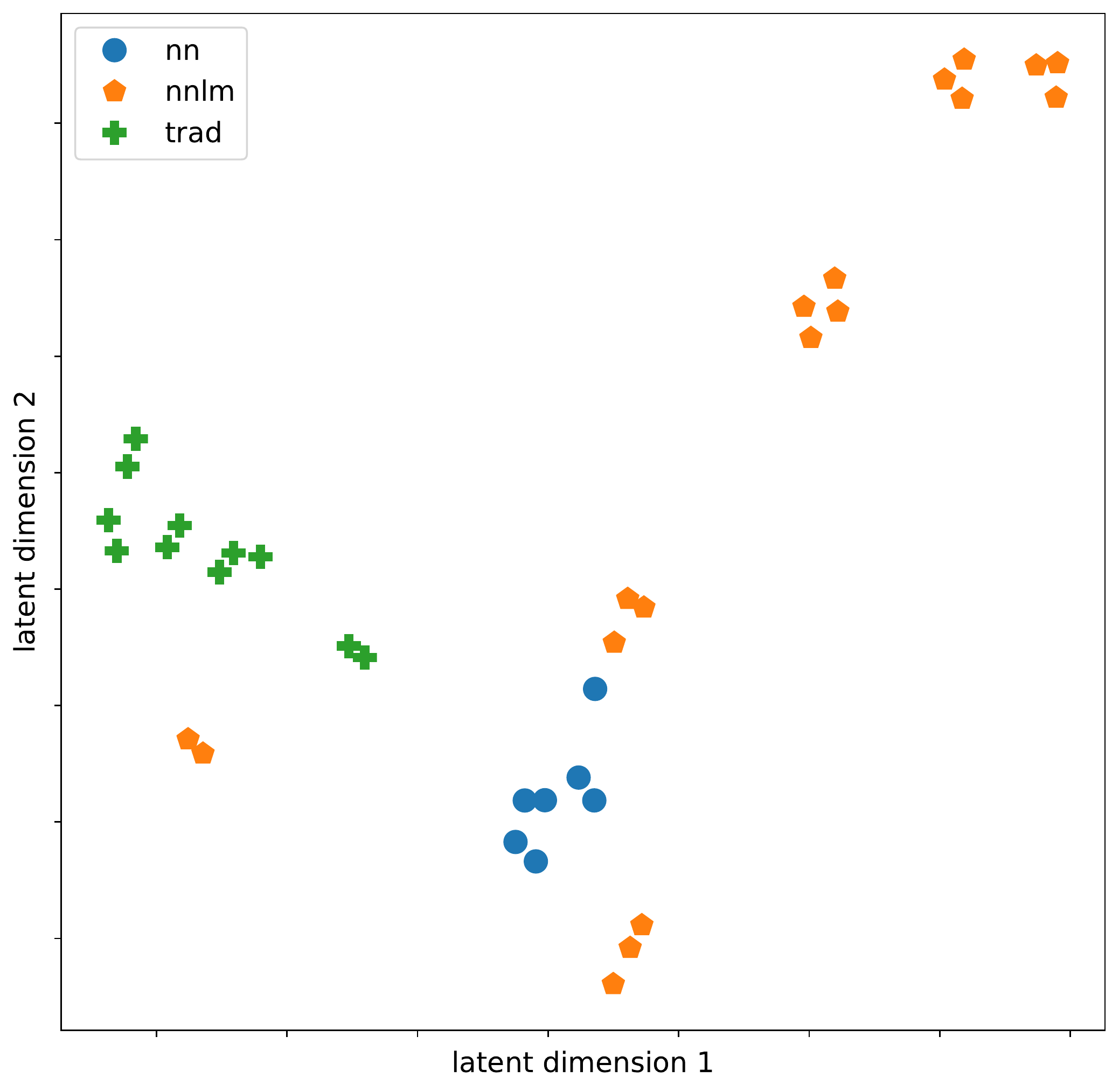}
    \caption{By model type on passage retrieval task}
    \label{fig:model-task-passages-by-model-type}
  \end{subfigure}
  \begin{subfigure}{.4\textwidth}
    \includegraphics[width=\textwidth]{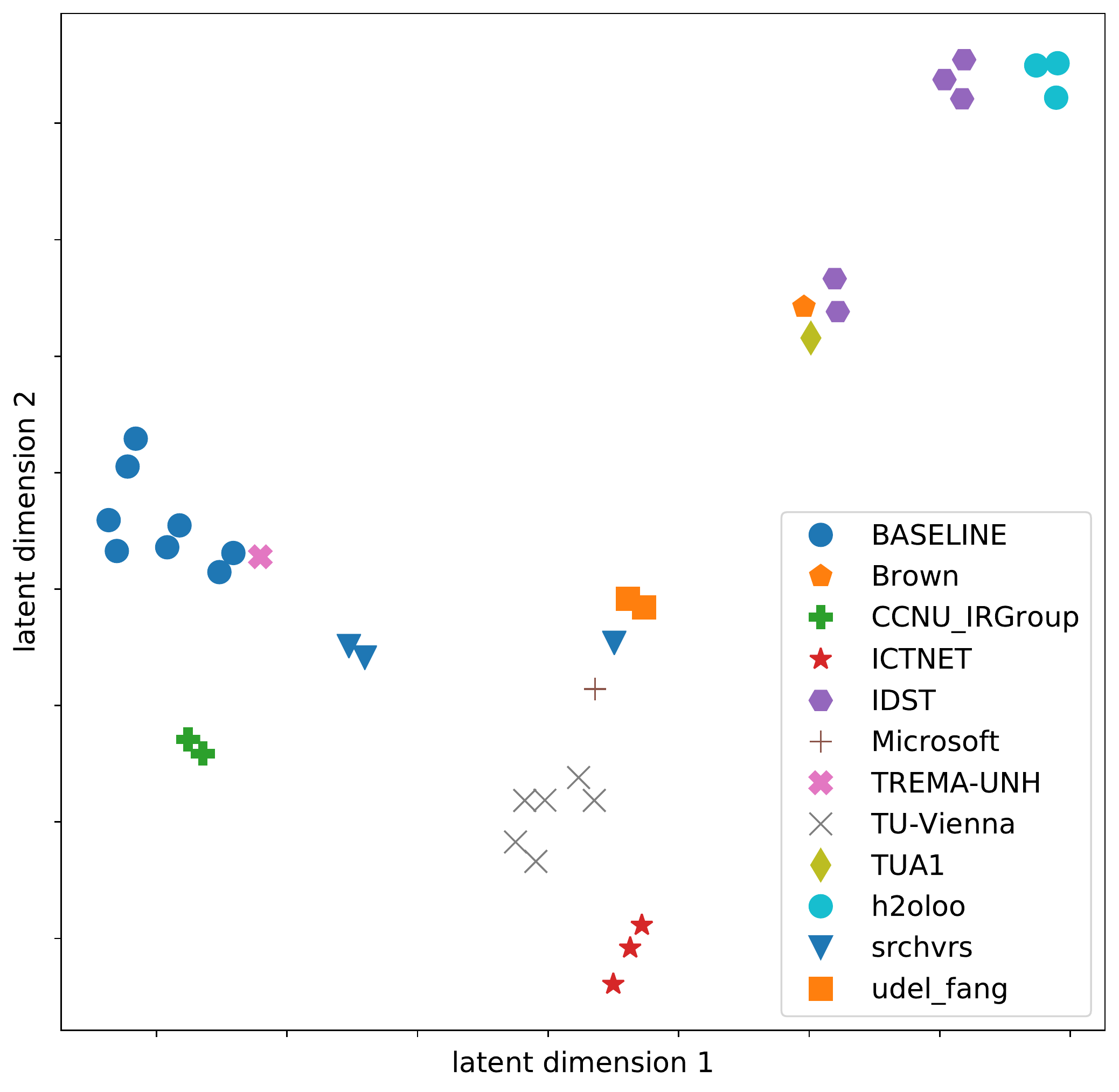}
    \caption{By group name on passage retrieval task}
    \label{fig:model-task-passages-by-team-name}
  \end{subfigure}
  \caption{Visualizing inter-run similarity using t-SNE.
  Each run is represented by a 43-dimensional vector of NDCG@10 performance on corresponding 43 test queries.
  The 43-dimensional vector is then reduced to two-dimensions and plotted using t-SNE.
  Runs that are submitted by the same group generally cluster together.
  Similarly, ``nnlm'', ``nn'', and ``trad'' runs also demonstrate similarities.}
  \label{fig:model}
\end{figure}

Next, we analyze this year's runs by representing each run as a vector of 43 NDCG@10 scores. In this vector space, two runs are similar if their NDCG vectors are similar, meaning they performed well and badly on the same queries. Using t-SNE \citep{maaten2008visualizing} we then plot the runs in two dimensions, which gives us a visualization where similar runs will be closer together and dissimilar results further apart. This method of visualizing inter-model similarity was first proposed by \citet{mitra2017learning} and we employ it to generate the plots in Figure~\ref{fig:model}.

On both document and passage retrieval tasks, the runs appear to be first clustered by group---see Figures~\ref{fig:model-task-docs-by-team-name} and \ref{fig:model-task-passages-by-team-name}.
This is expected, as different runs from the same group are likely to employ variations of the same approach.
In Figures~\ref{fig:model-task-docs-by-model-type} and \ref{fig:model-task-passages-by-model-type}, runs also cluster together based on their categorization as ``nnlm'', ``nn'', and ``trad''.

\paragraph{End-to-end retrieval \vs reranking.}


Our datasets include top-$k$ candidate result lists, with 100 candidates per query for document retrieval and 1000 candidates per query for passage retrieval. Runs that simply rerank the provided candidates are ``rerank'' runs, whereas runs that perform end-to-end retrieval against the corpus, with millions of potential results, are ``fullrank'' runs. We would expect that a ``fullrank'' run should be able to find a greater number of relevant candidates than we provided, achieving higher NCG@$k$. A multi-stage ``fullrank'' run should also be able to optimize the stages jointly, such that early stages produce candidates that later stages are good at handling.

\begin{figure}
  \center
  \begin{subfigure}{.49\textwidth}
    \includegraphics[width=\textwidth]{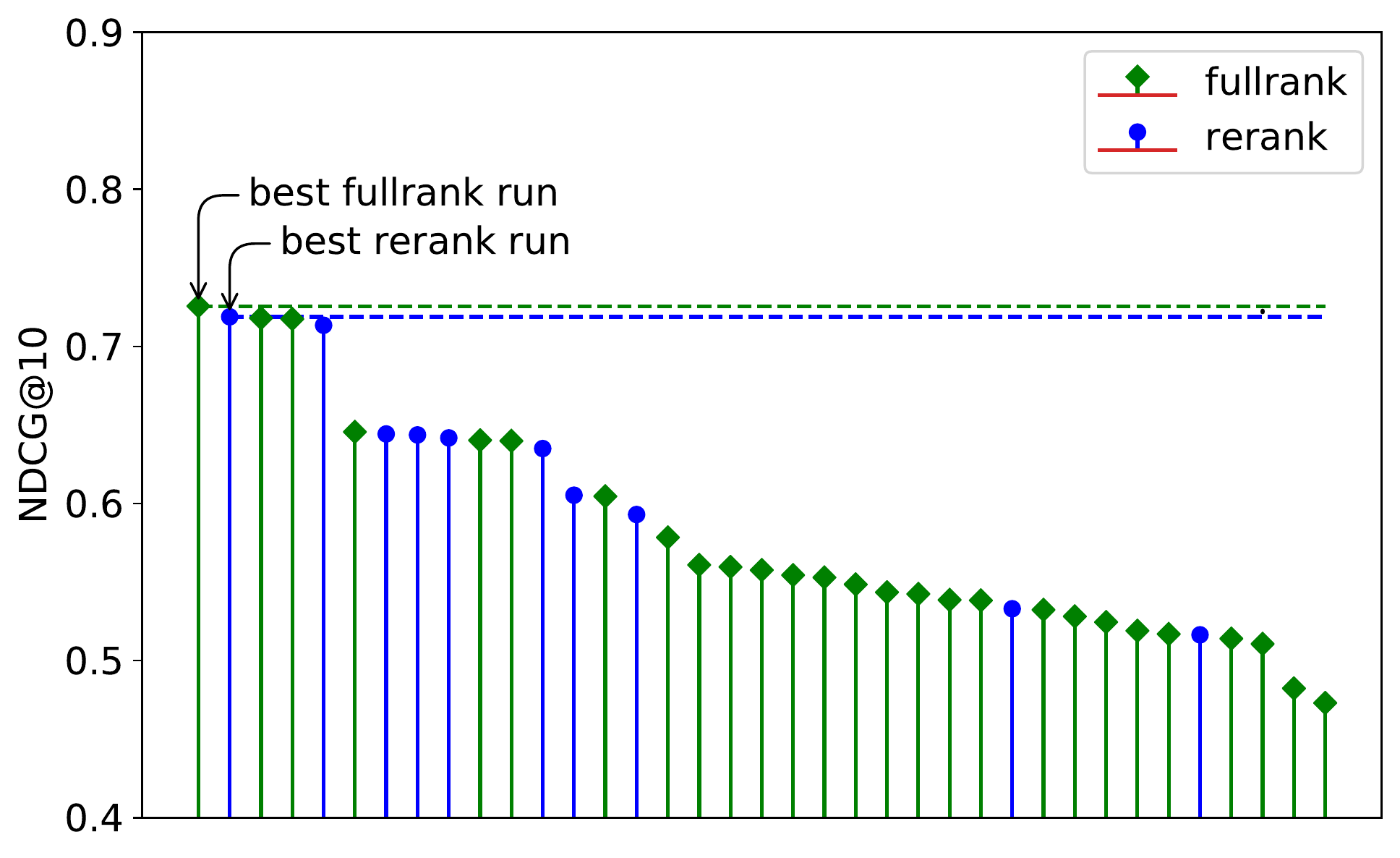}
    \caption{NDCG@10 for runs on the document retrieval task}
    \label{fig:model-task-docs-stem-by-subtask}
  \end{subfigure}
  \hfill
  \begin{subfigure}{.49\textwidth}
    \includegraphics[width=\textwidth]{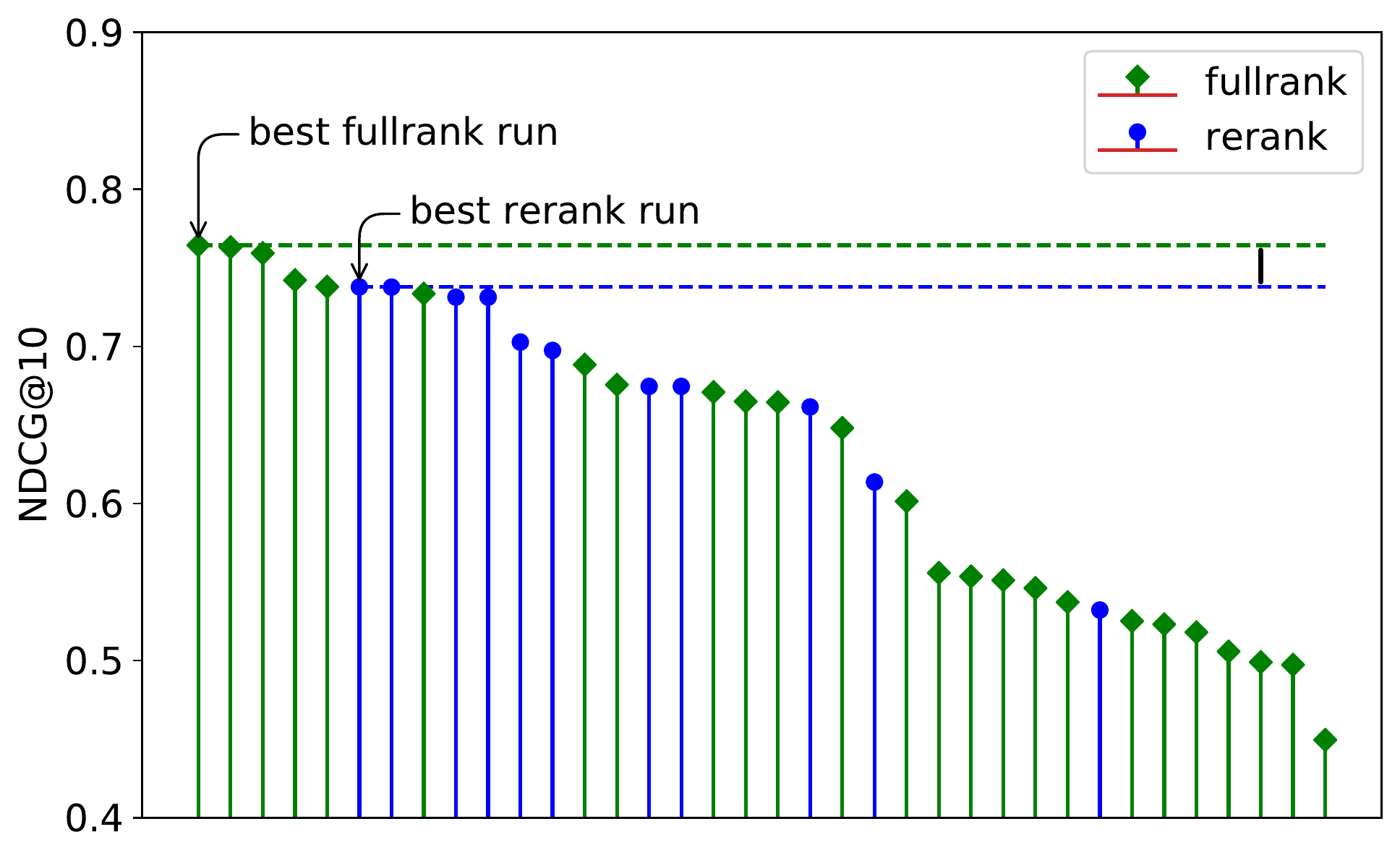}
    \caption{NDCG@10 for runs on the passage retrieval task}
    \label{fig:model-task-passages-stem-by-subtask}
  \end{subfigure}
  \begin{subfigure}{.49\textwidth}
    \includegraphics[width=\textwidth]{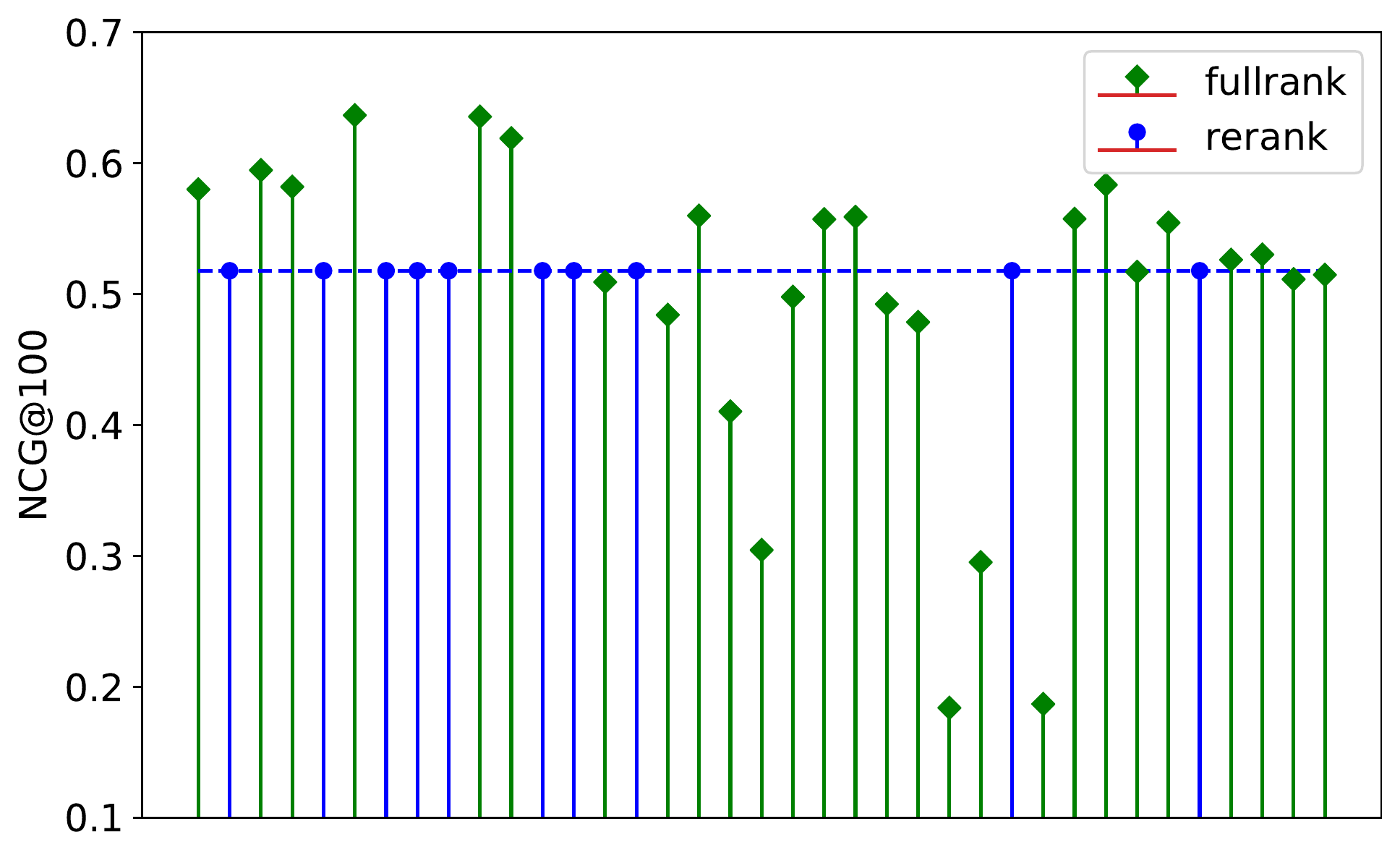}
    \caption{NCG@100 for runs on the document retrieval task}
    \label{fig:recall-task-docs-stem}
  \end{subfigure}
  \hfill
  \begin{subfigure}{.49\textwidth}
    \includegraphics[width=\textwidth]{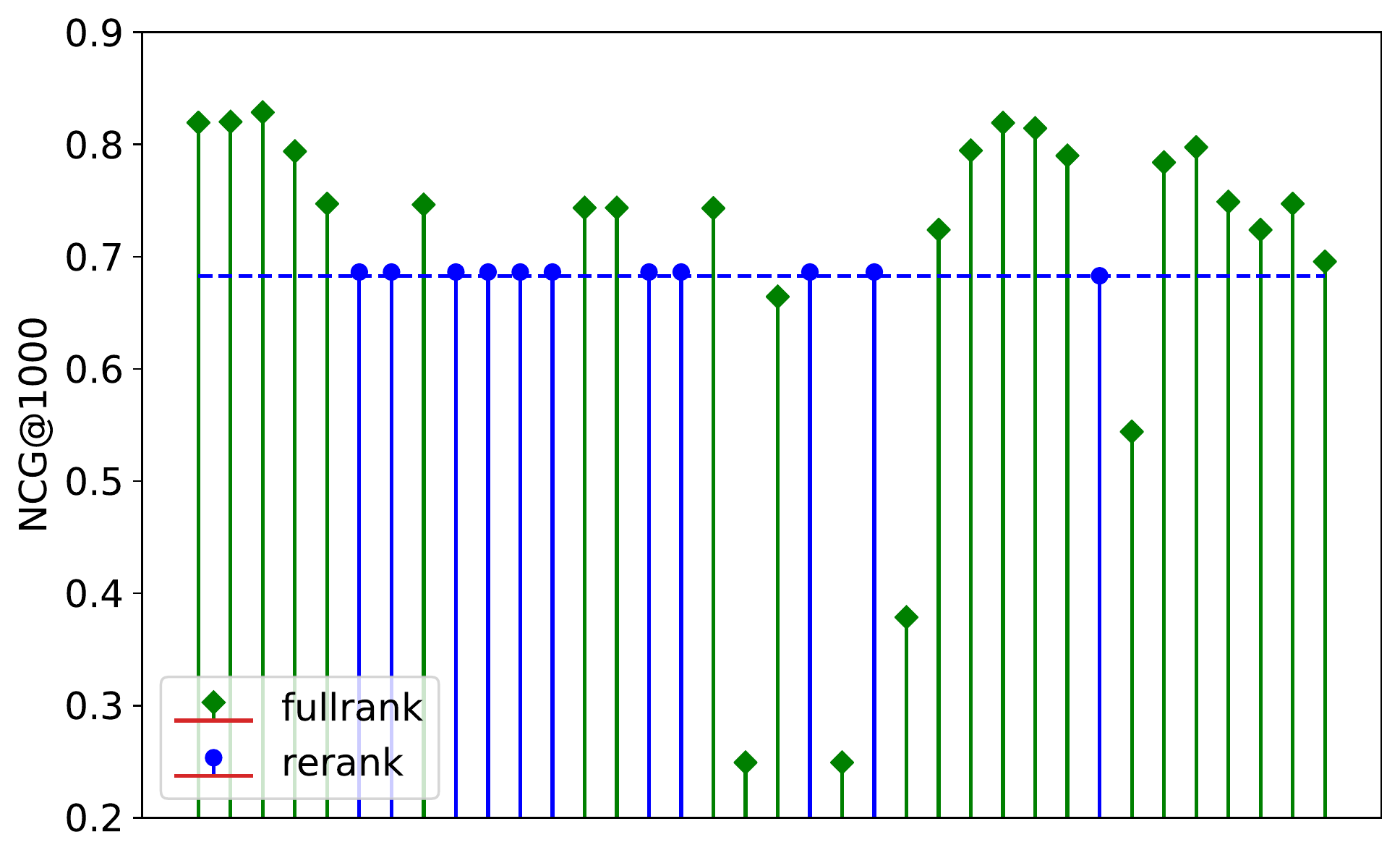}
    \caption{NCG@1000 for runs on the passage retrieval task}
    \label{fig:recall-task-passages-stem}
  \end{subfigure}
  \caption{Analyzing the impact of ``fullrank'' \vs ``rerank'' settings on retrieval performance.
  Figure~(a) and (b) show the performance of different runs on the document and passage retrieval tasks, respectively.
  Figure~(c) and (d) plot the NCG@100 and NCG@1000 metrics for the same runs for the two tasks, respectively.
  The runs are ordered by their NDCG@10 performance along the $x$-axis in all four plots.
  We observe, that the best run under the ``fullrank'' setting outperforms the same under the ``rerank'' setting for both document and passage retrieval tasks---although the gaps are relatively smaller compared to those in Figure~\ref{fig:model-stem-by-model-type}.
  If we compare Figure~(a) with (c) and Figure~(b) with (d), we do not observe any evidence that the NCG metric is a good predictor of NDCG@10 performance.}
  \label{fig:recall-stem}
\end{figure}

According to Figure~\ref{fig:recall-stem}, ``fullrank'' did not achieve much better NDCG@10 performance than ``rerank'' runs. While it was possible for ``fullrank'' to achieve better NCG@$k$, it was also possible to make NCG@$k$ worse, and achieving significantly higher NCG@$k$ does not seem necessary to achieve good NDCG@10. 

Specifically, for the document retrieval task, the best ``fullrank'' run achieves only $0.9\%$ higher NDCG@10 over the best ``rerank' run.
For the passage retrieval task, the difference is $3.6\%$.

The best NCG@100 for the document retrieval task is achieved by a well-tuned combination of BM25 \citep{robertson2009probabilistic} and RM3 \citep{abdul2004umass} on top of document expansion using doc2query \citep{nogueira2019document}---which improves by $22.9\%$ on the metric relative to the set of $100$ candidates provided for the reranking task.
For the passage retrieval task, the best NCG@1000 is $20.7\%$ higher than that of the provided reranking candidate set.

Given this was the first ever Deep Learning Track at TREC, we are not yet seeing a strong advantage of ``fullrank'' over ``rerank''.
However, we hope that as the body of literature on neural methods for phase 1 retrieval (\eg, \citep{boytsov2016off, zamani2018neural, mitra2019incorporating, nogueira2019document}) grows, we would see a larger number of runs with deep learning as an ingredient for phase 1 in future editions of this TREC track.

\paragraph{NIST labels \vs Sparse MS MARCO labels.}

\begin{figure}
    \centering
    \begin{subfigure}{0.7\textwidth}
    \includegraphics[width=\textwidth]{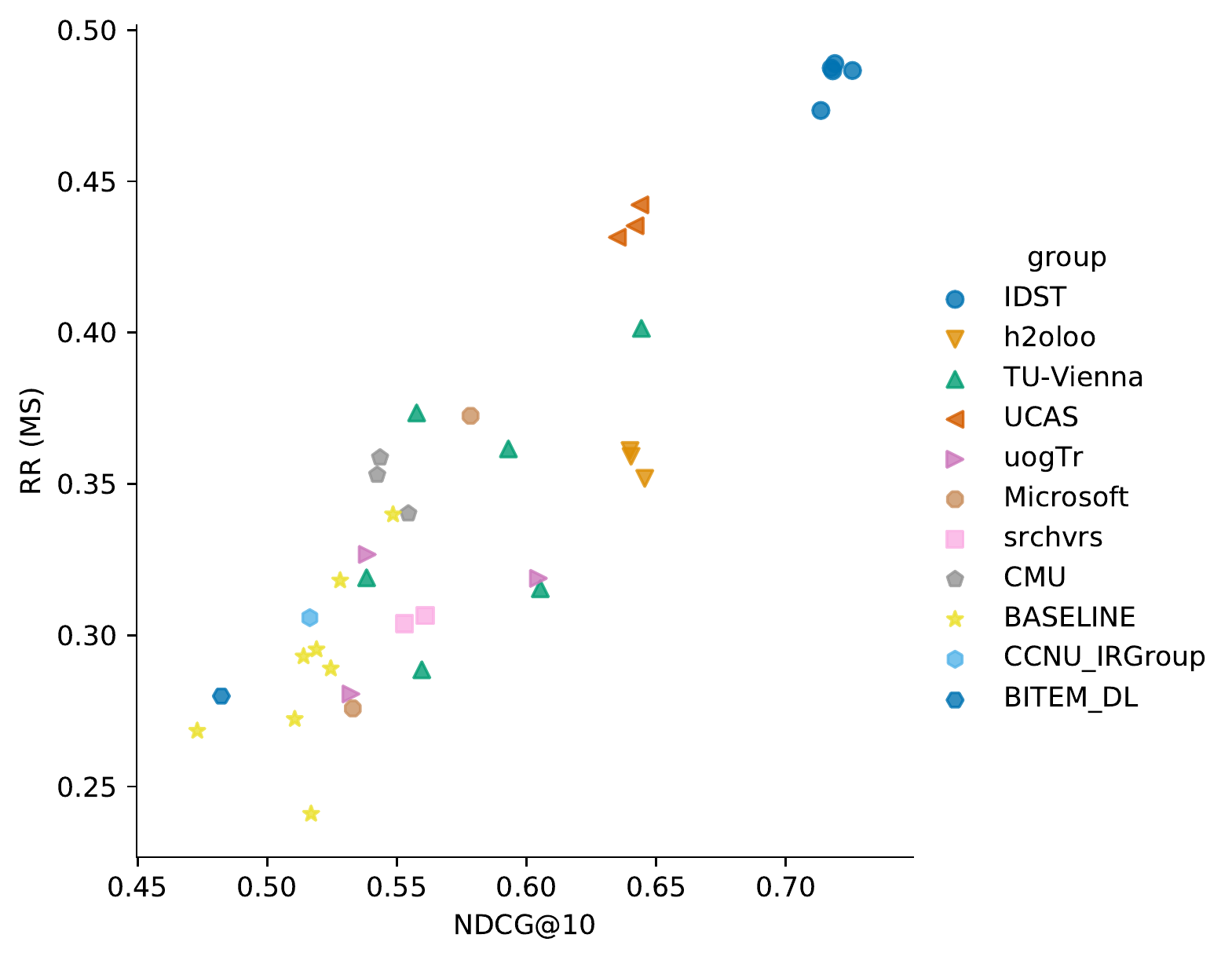}
    \caption{Document retrieval task.}
    \end{subfigure}
    \begin{subfigure}{0.7\textwidth}
    \includegraphics[width=\textwidth]{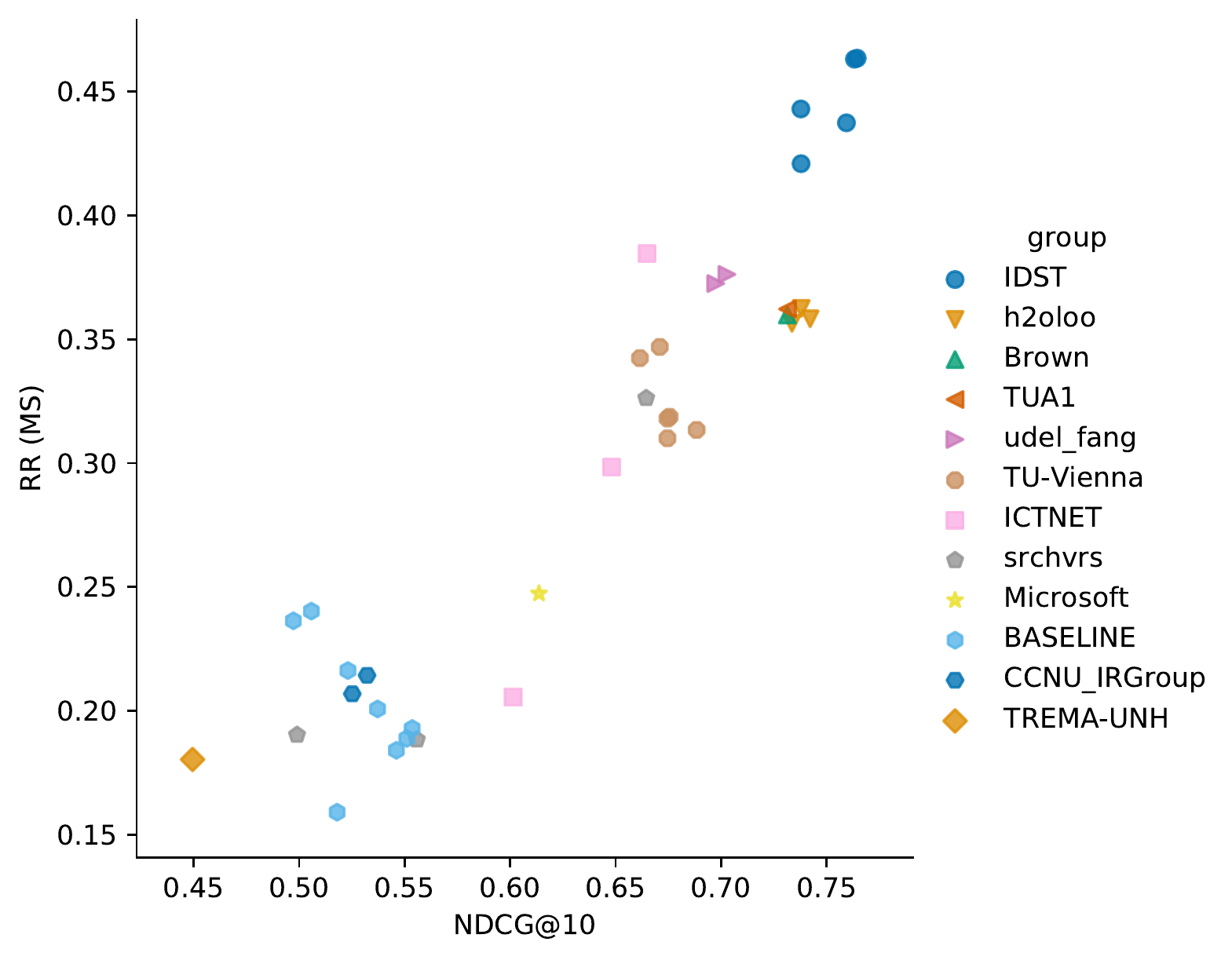}
    \caption{Passage retrieval task.}
    \end{subfigure}
    \caption{Metrics agreement scatter plot, broken down by group. RR (MS) is reciprocal rank calculated with the sparse MS MARCO labels, while NDCG@10 is calculated using NIST labels.}
    \label{fig:my_label}
\end{figure}


\begin{figure}
    \centering
    \includegraphics[width=\textwidth]{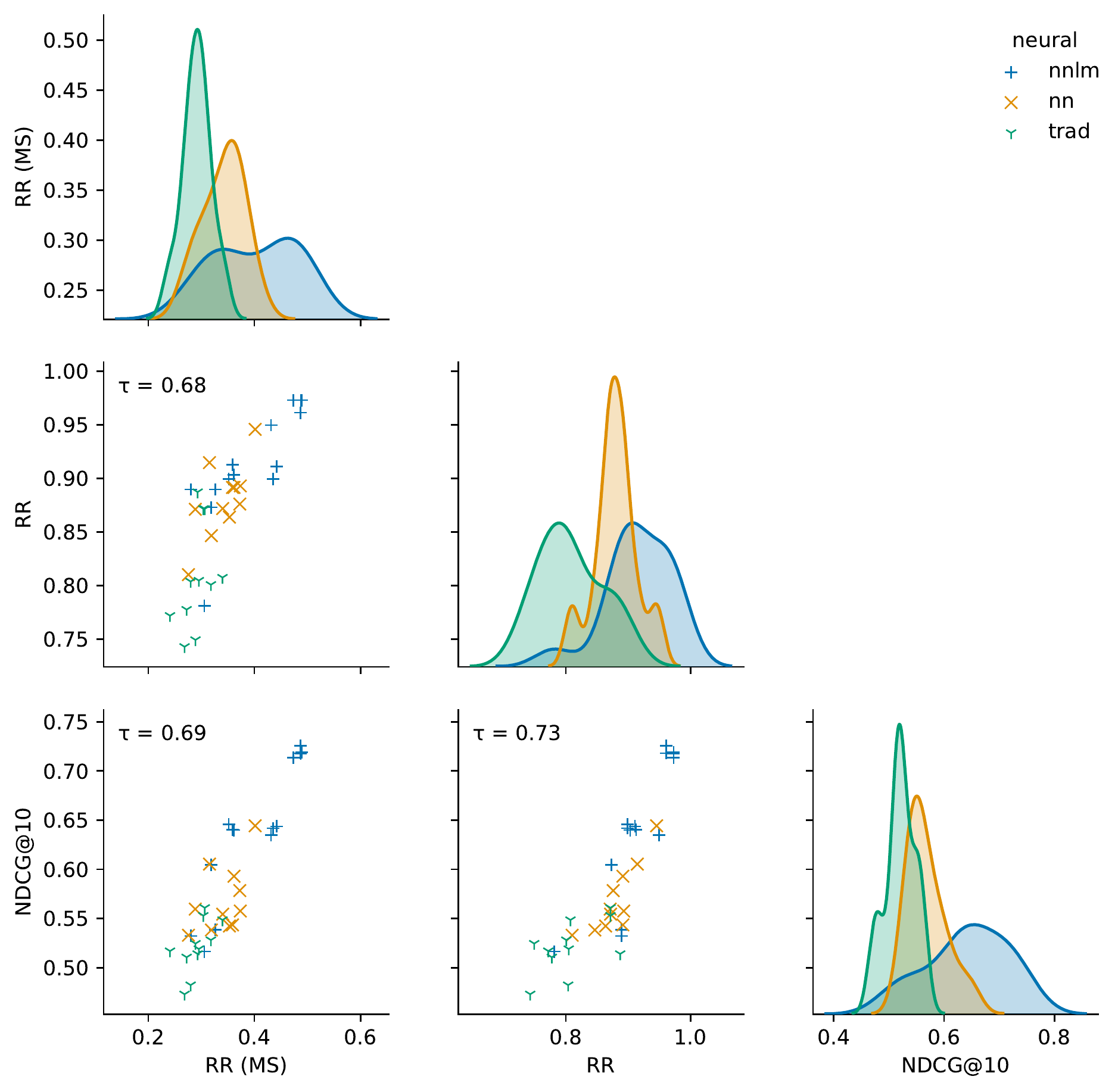}
    \caption{Metrics agreement analysis, broken down by model type, for the document retrieval task. Kendall correlation ($\tau$) indicates agreement between metrics on system ordering. RR (MS) is calculated using MS MARCO sparse labels, while RR and NDCG@10 are calculated using NIST labels.}
    \label{fig:metric-task-docs-vs}
\end{figure}

\begin{figure}
    \centering
    \includegraphics[width=\textwidth]{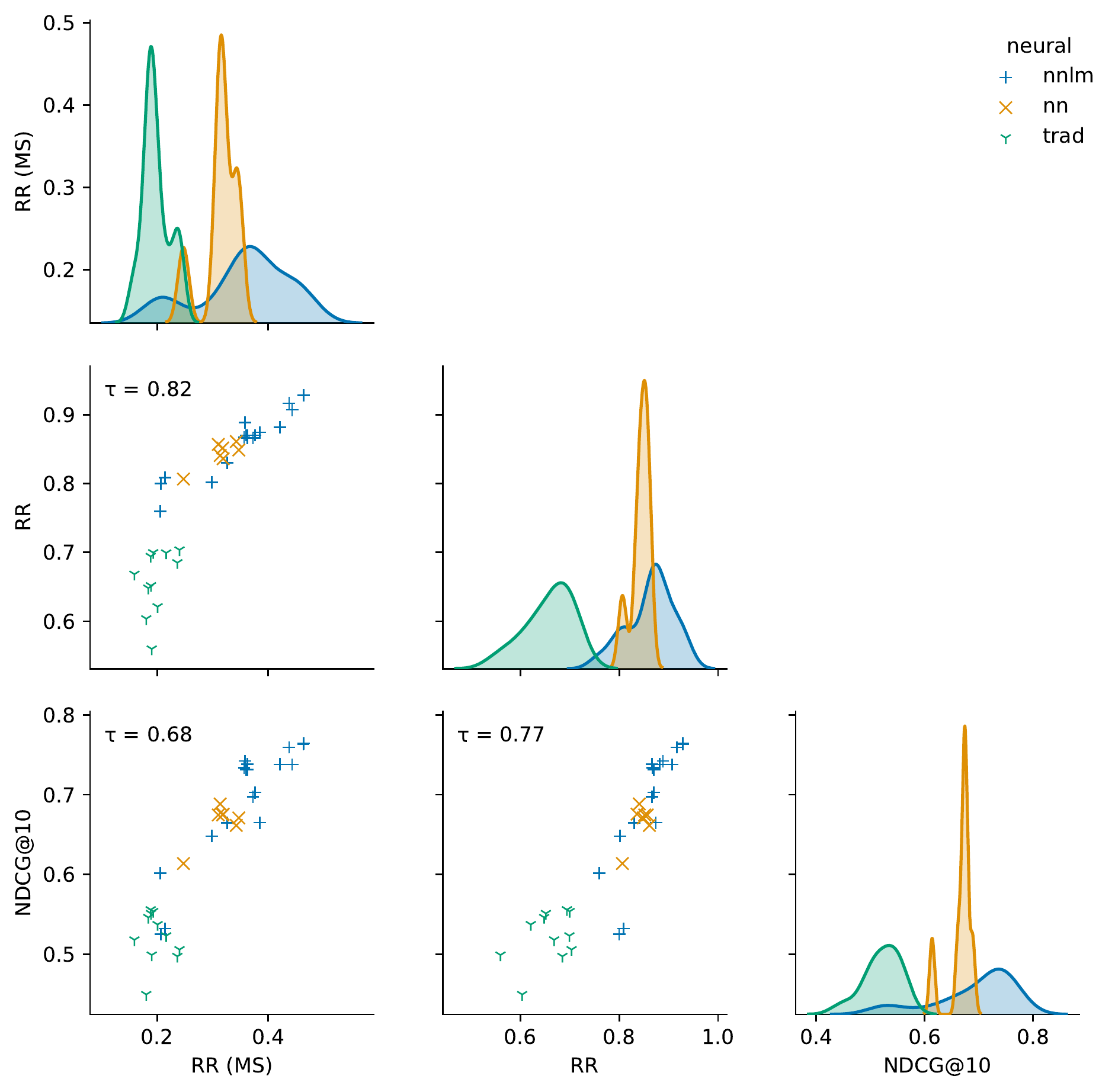}
    \caption{Metrics agreement analysis, broken down by model type, for the passage retrieval task. Kendall correlation ($\tau$) indicates agreement between metrics on system ordering. RR (MS) is calculated using MS MARCO sparse labels, while RR and NDCG@10 are calculated using NIST labels.}
    \label{fig:metric-task-passages-vs}
\end{figure}

Our baseline human labels from MS MARCO often have one known positive result per query. We use these labels for training, but they are also available for test queries. Although our official evaluation uses NDCG@10 with NIST labels, we now compare this with reciprocal rank (RR) using MS MARCO labels, and RR using NIST labels. Our goal is to understand how changing the labeling scheme and metric affects the overall results of the track, but if there is any disagreement we believe the NDCG results are more valid, since they evaluate the ranking more comprehensively and a ranker that can only perform well on labels with exactly the same distribution as the training set is not robust enough for use in real-world applications, where real users will have opinions that are not necessarily identical to the preferences encoded in sparse training labels.





In Figure~\ref{fig:metric-task-docs-vs} and \ref{fig:metric-task-passages-vs}, We observe general agreement between results using MS MARCO and NIST labels--\ie, runs that perform well on MS MARCO-style evaluation also tends to achieve good performance when evaluated under traditional TREC settings, and vice versa.
This is good news, validating the MS MARCO leaderboard results are at least somewhat indicative of results that are found with pooled judging.



\section{Reusability of test collections}
\label{sec:reusability}

One goal of the track was to create traditional ad~hoc test sets 
based on the MS MARCO dataset within available budgets.
Since the Document Ranking and Passage Ranking tasks used different document
sets, two separate test collections, one per task, were constructed.
The two test collections started from a common set of topics and each topic was judged by
the same NIST assessor for both documents and passages, but
assessing for documents and passages was done at  different times.
Further, the evaluation set of topics (i.e., the topics over which evaluation scores
are computed) are overlapping but not identical in the two collections.
Thus the collections created in the track are two separate, independent collections.

The runs submitted to the track consisted of ranked lists of items for each topic in
the test set of 200 topics.  NIST selected 52 topics from this set to be judged.
The topics were selected by observing the behavior of submitted Document Ranking
task runs on the entire test set when using the sparse MARCO judgments to evaluate runs.
Test questions that had median MRR scores greater than 0.0 but no more than 0.5 were candidates to be judged.

The judgment process then proceeded as follows, where the items to be judged will
generically be called `documents' even though those documents were MS MARCO passages
for the Passage Ranking task.
\begin{enumerate}
\item For each question, create a top-10 pool across all runs in the task,
      and add any document that contains a judgment in the MARCO sparse judgments.
      Call the size of this set $P$ (which varies from topic to topic). 
      The assessor judges these pool documents first, then another 100 documents
      selected to be judged using the University of Waterloo's HiCAL~\citep{HiCAL} system.
      HiCAL uses the current set of judgments to build a relevance model and then selects
      the unjudged document most likely to be relevant as the next document to judge.
      At the end of this stage there are $R$ known relevant documents.
      If $2R < P$, the judging is finished for this topic.
\item Call the the difference between the number of documents that have been judged
      and the desired number of $2R+100$ judgments $G$.
      Judge another $G$ documents selected by HiCAL.
      Now the number of judgments for the topic is $J = P+100+G$
      and the new number of known relevant is $R^*$. If $2R^*+100 < J$,
      assessment is finished for the topic.
      If $R^* \approx J$, then discard the topic because it will be
      too expensive to get ``sufficiently complete'' judgments for it.
\item If a topic is still live, add a new increment proportional to the number of
      known relevant documents to the topic budget, and iterate, terminating when
      (if) the number of known relevant documents is less than half the number of
      judged documents.
\item Terminate the entire process when assessors are out of time or have nothing
      left to judge.
\end{enumerate}
The resulting evaluation set was the set of topics with
at least three relevant documents and a ratio of $R^* / J < 0.6$.
This process resulted in 43 topics in the evaluation set for both the Document Ranking
and the Passage Ranking tasks, but as noted it is a slightly different 43 topics
for the two tasks.

Documents in the Document Ranking task were judged on a four-point scale
of {\em Irrelevant}~(0), {\em Relevant}~(1), {\em Highly Relevant}~(2),
and {\em Perfectly Relevant}~(3) where all but Irrelevant were treated as
relevant in HiCAL and in computing binary-relevance-based measures.
For the Passage Ranking task, passages were judged on a four-point scale
of {\em Irrelevant}~(0), {\em Related} (the passage is on-topic but does not answer the question)~(1),
{\em Highly Relevant}~(2),  and  {\em Perfectly Relevant}~(3).
In this task, only Highly and Perfectly Relevant were considered to be relevant
for binary measures and by HiCAL, though nDCG scores did use a gain value of 1
for the Related passages.
Table~\ref{tbl:rels} gives counts of the number of documents judged and the number of
relevant documents (using the definitions for binary relevance) found for each
of the 52 topics that entered the process.
\begin{table*}
\begin{center}
\caption{Judging statistics for the Document Ranking and Passage Ranking tasks.
Given are the number of documents judged (any variant of) relevant, the total
number of  documents judged, and the fraction of judged documents that are
relevant (Relevant Ratio).
Topics were excluded from the evaluation set if they had fewer than 3 relevant 
or if the fraction of judged documents that are relevant was greater than 0.6.
Data for excluded topics are given \textcolor{gray}{in gray}.
The final rows gives the total number of documents judged
and the number of documents judged when not counting excluded topics.}

\footnotesize

\begin{tabular}{ccrrrcrrr}
  & \hspace{5em} &
  \multicolumn{3}{c}{Document Ranking} & \hspace{3em} &
  \multicolumn{3}{c}{Passage Ranking} \\
\multicolumn{1}{c}{Topic} & &
	\multicolumn{1}{c}{\# Relevant} & \multicolumn{1}{c}{\# Judged} &
 		\multicolumn{1}{c}{Relevant Ratio} &  &
	\multicolumn{1}{c}{\# Relevant} & \multicolumn{1}{c}{\# Judged} &
		\multicolumn{1}{c}{Relevant Ratio} \\ 
\cline{1-1} \cline{3-5} \cline{7-9}
19335 & & 53  & 239 & 0.222  & &  7 & 194 & 0.036 \\
47923 & &  767  &  1476 & 0.520 & & 41 & 143 & 0.287 \\
87181 & & 168  &  404 & 0.416 & & 31 & 158 & 0.196 \\
87452 & & 165  &  346 & 0.477 & & 31 & 139 & 0.223 \\
100983 & & \textcolor{gray}{341}  &  \textcolor{gray}{420} & \textcolor{gray}{0.812} &
           & \textcolor{gray}{370} & \textcolor{gray}{432} & \textcolor{gray}{0.856} \\
104861 & & 61  &  218 & 0.280 & & 111 & 306 & 0.363 \\
130510 & & 42  &  174 & 0.241 & & 14 & 133 & 0.105 \\
131843 & & 25  &  168 & 0.149 & & 19 & 132 & 0.144 \\
146187 & & 25  &  157 & 0.159 & & 8 & 138 & 0.058 \\
148538 & & 240  &  578 & 0.415  & & 32 & 159 & 0.201 \\
156493 & & 151  &  378 & 0.399 & & 117 & 300 & 0.390 \\
168216 & & \textcolor{gray}{578}  &  \textcolor{gray}{885} & \textcolor{gray}{0.653} &
          & 200 & 582 & 0.344 \\
182539 & &  23  &  144 & 0.160 & & 9 & 132 & 0.068 \\
183378 & &  324  &  723 & 0.448 & & 175 & 451 & 0.388 \\
207786 & &  76  &  228 & 0.333 & & 11 & 137 & 0.080 \\
264014 & & 177  &  415 & 0.427 & & 152 & 382 & 0.398 \\
287683 & &  3  &  190 & 0.016 & 
          & \textcolor{gray}{1} & \textcolor{gray}{140} & \textcolor{gray}{0.007} \\
359349 & &  183  &  446 & 0.410 & & 25 & 139 & 0.180 \\
405717 & &  34  &  171 & 0.199 & & 7 & 144 & 0.049 \\
423273 & &  \textcolor{gray}{1}  &  \textcolor{gray}{183} & \textcolor{gray}{0.005} &
          & \textcolor{gray}{2} & \textcolor{gray}{199} & \textcolor{gray}{0.010} \\
443396 & &  195  &  376 & 0.519 & & 63 & 188 & 0.335 \\
451602 & & 202  &  415 & 0.487 &  & 100 & 220 & 0.455 \\
489204 & &  392  &  700 & 0.560 & & 24 & 175 & 0.137 \\
490595 & &  51  &  161 & 0.317 & & 24 & 148 & 0.162 \\
527433 & &  52  &  204 & 0.255 & & 34 & 160 & 0.212 \\
573724 & &  42  &  176 & 0.239 & & 13 & 141 & 0.092 \\
833860 & &  178  &  412 & 0.432 & & 42 & 157 & 0.268 \\
855410 & &  5  &  337 & 0.015 & & 3 & 183 & 0.016 \\
915593 & &  115  &  314 & 0.366 & & 79 & 192 & 0.411 \\
962179 & &  24  &  173 & 0.139 & & 21 & 161 & 0.130 \\
966413 & &  \textcolor{gray}{283}  &  \textcolor{gray}{372} & \textcolor{gray}{0.761} &
          & \textcolor{gray}{120} & \textcolor{gray}{180} & \textcolor{gray}{0.667} \\
1037798 & &  44  &  188 & 0.234 & & 7 & 154 & 0.045 \\
1063750 & &  381  &  708 & 0.538 & & 183 & 392 & 0.467 \\
1103812 & &  40  &  234 & 0.171 & & 11 & 141 & 0.078 \\
1104031 & &  \textcolor{gray}{432} &  \textcolor{gray}{466} & \textcolor{gray}{0.927} &
           & \textcolor{gray}{113} & \textcolor{gray}{152} & \textcolor{gray}{0.743} \\
1104492 & &  \textcolor{gray}{335}  &  \textcolor{gray}{395} & \textcolor{gray}{0.848} &
           & \textcolor{gray}{192} & \textcolor{gray}{300} & \textcolor{gray}{0.640} \\
1106007 & &  242  &  416 & 0.582 & & 41 & 178 & 0.230 \\
1110199 & &  41  &  183 & 0.224 & & 28 & 175 & 0.160 \\
1112341 & &  385  &  664 & 0.580 & & 119 & 223 & 0.534 \\
1113437 & &  93  &  280 & 0.332 & & 25 & 180 & 0.139 \\
1114646 & &  55  &  163 & 0.337 & & 12 & 151 & 0.079 \\
1114819 & &  562  &  1026 & 0.548 & & 213 & 470 & 0.453 \\
1115776 & &  7  &  158 & 0.044 & & 4 & 152 & 0.026 \\
1117099 & &  386  &  845 & 0.457 & & 83 & 257 & 0.323 \\
1121402 & &  55  &  200 & 0.275 & & 23 & 146 & 0.158 \\
1121709 & &  \textcolor{gray}{2}  &  \textcolor{gray}{250} & \textcolor{gray}{0.008} &
	   & 3 & 178 & 0.017 \\
1121986 & &  \textcolor{gray}{440}  &  \textcolor{gray}{474} & \textcolor{gray}{0.928} &
           & \textcolor{gray}{263} & \textcolor{gray}{378} & \textcolor{gray}{0.696} \\
1124210 & &  276  &  629 & 0.439 & & 120 & 330 & 0.364 \\
1129237 & &  38  &  175 & 0.217 & & 17 & 147 & 0.116 \\
1132213 & &  20  &  204 & 0.098 &
          & \textcolor{gray}{0} & \textcolor{gray}{163} & \textcolor{gray}{0.000} \\
1133167 & &  199  &  464 & 0.429 & & 219 & 492 & 0.445 \\
1134787 & &  \textcolor{gray}{426}  & \textcolor{gray}{454} & \textcolor{gray}{0.938} &
	   & \textcolor{gray}{467} & \textcolor{gray}{700} & \textcolor{gray}{0.667}\\
\hline \hline
\multicolumn{2}{l}{Total judged:} & \multicolumn{3}{r}{20,157} &
		& \multicolumn{3}{r}{11,904}\\
\multicolumn{2}{l}{Final qrels size:} & \multicolumn{3}{r}{16,258} &
		& \multicolumn{3}{r}{9260}\\
\end{tabular}
\label{tbl:rels}
\end{center}
\end{table*}

HiCAL is a dynamic collection construction method, meaning that the document
to be judged next is selected only after judgments for previous documents
have been received.
The Common Core track in TRECs~2017 and 2018 used a method based on multi-armed
bandit optimization techniques, another dynamic method, with the similar goal
of building high-quality, reusable, ad~hoc test collections affordably~\citep{CIKMbandits}.
That work showed two main issues to be overcome when building new collections
with dynamic techniques: providing the assessors the opportunity to learn
a topic before immutable judgments are rendered, and setting individual topic
budgets when assessors judge at different rates and at different times but are
subject to a single overall judgment budget.
The first issue is less severe with (NIST's modification of) HiCAL since assessors
can change the value of any previously made judgment at any time; whenever a new
relevance model is calculated, HiCAL uses the judgments current at the time of calculation.
Nonetheless, top-10 pools provide both an opportunity for assessors to learn a topic
and ensure that all measures based on document-level cutoffs less
than or equal to ten are precise for all judged runs,
and this motivated the use of pools in the first stage of the process.
Setting per-topic judgment budgets continues to be a challenging problem.
The stopping criterion of ending a topic once $2R + 100$ documents were judged
was motivated by the heuristic observed by Waterloo in prior use
of HiCAL~\citep{BeyondPooling}
further supported by the Common Core track's observation that a topic 
for which more than half of its judged documents are relevant is unlikely
to be sufficiently judged\footnote{We nonetheless included topics with a ratio
of relevant to judged between 0.5 and 0.6 in the evaluation set because test
collection stability tests suggest the collection is more stable with those
topics than without them (likely because the total number of topics is greater
with them) and to provide a greater diversity of topic sizes in the evaluation set.}.
Note that the process described above was the target process, but
the practicalities of keeping assessors occupied meant that some topics
received more judgments than they ``deserved''.
All judgments for non-excluded topics are included in the qrels file.

\subsection{Collection Robustness}
Our goal is to build general-purpose, reusable test collections at acceptable cost.
In this context, general-purpose means a collection reliably ranks runs
for a wide spectrum of evaluation measures, including recall-focused measures.
Reusable means that runs that did not participate in the collection building process
can be reliably ranked by the collection.
Since costs in building a collection are generally dominated by the cost
of human assessments, the number of relevance
judgments required is used as the construction cost. 

Leave-Out-Uniques (LOU) tests~\citep{IRPooling,zob:sigir98} are a way of analyzing
the reusability of a collection.
In these tests, the relevant documents retrieved by only one participating team
are removed from the qrels files and all runs are then evaluated using the
reduced qrels.
The reduced qrels are the qrels that would have resulted had the team
not participated in the collection building process, and thus their submitted runs
represent  new runs with respect to the reduced qrels.
If the ranking of runs using the reduced qrels is essentially the same as the
ranking of runs using the original qrels over all participating teams, then the
original collection is likely reusable.
The similarity between rankings of runs is usually defined by the Kendall's $\tau$
correlation between the rankings. Kendall's $\tau$ is a measure of association that
is proportional to the number of interchanges between adjacent items
in one ranking that are required to turn that ranking into the other.
$\tau$ scores are normalized such that a score of 1 designates perfect agreement,
-1 designates rankings that are inverses of one another, and 0 designates rankings
that are independent of one another.
$\tau$ scores can be misleading in the case of system rankings of TREC submissions,
however, because usually there are a set of very good runs and a set of very poor
runs and each of those run sets always rank in the same order.
Thus, in addition to the $\tau$ score between the rankings, we also report drops,
the largest (negative) difference in ranks experienced by some run~\citep{CIKMbandits}.

A standard LOU test does not work for examining the collections built in
the Deep Learning track because the HiCAL process does not depend on
runs to provide documents and thus ``unique relevant documents'' is no
longer a well-defined concept.
A given team's unique relevant documents can be removed from the depth-10 pools
in the first stage, but then the HiCAL process must activated as it may select
the removed documents to be judged in later stages.
Since the HiCAL process is not deterministic (ties are broken randomly) and
depends on the particular set of documents seen so far, the HiCAL process
must be simulated multiple times using the original qrels' judgments.

The simulations proceeded as follows, where the entire process was performed
separately for the Document Ranking and Passage Ranking collections.
The original depth-10 pools (i.e., top-10 documents from all runs plus
MARCO judgments) were fed to the HiCAL process for each of ten trials, where each
trial used a separate initial seed for the random number generator.
Within each trial, we tracked the documents encountered by HiCAL, creating a trace
of the first 2500 documents encountered per topic.
Any unjudged documents encountered by HiCAL were treated as not relevant.
We created a qrels file from each trace by taking a prefix of the trace of length
equal to the number of documents judged in the original qrels per topic.
This resulted in 10 qrels files that could have resulted as the official qrels of
the track (modulo the unjudged documents would have been judged).
While these qrels are not identical to one another nor to the official qrels,
they do rank systems very similarly.
The leftmost segment of Table~\ref{tau.tab} shows the $\tau$ values and the drops for MAP
scores over the set of ten trials\footnote{Prec(10) scores are identical over all trials
because each trial starts with a depth-10 pool.}.  The top part of the table gives statistics
for the Document Ranking task collection and the bottom part for the Passage Ranking task
collection.
\begin{table*}
\begin{center}
\caption{Kendall's $\tau$ and Maximum Drop in ranks observed in simulation trials.
Each trial creates a qrels file of the same size as the official qrels, and the ranking
of systems induced by that qrels is compared to the ranking induced by the official
qrels. Using all team's runs compared to the original (left columns) shows the effect of the
nondeterminism of HiCAL. The remainder of the columns show the effect of omitting 
one team's runs from the pools in the first stage.}
\begin{tabular}{crccrcrc}
 & \multicolumn{2}{c}{All vs. Official} & \hspace{.5in} &
		\multicolumn{4}{c}{Omit Team vs. Official}\\
 & \multicolumn{2}{c}{MAP} & & \multicolumn{2}{c}{MAP} & \multicolumn{2}{c}{Prec(10)} \\
\multicolumn{1}{c}{Trial} & \multicolumn{1}{c}{$\tau$} & \multicolumn{1}{c}{Drop} & &
	\multicolumn{1}{c}{$\tau$} & \multicolumn{1}{c}{Drop} &
	\multicolumn{1}{c}{$\tau$} & \multicolumn{1}{c}{Drop} \\
\cline{2-3} \cline{5-6} \cline{7-8}
1 & 0.9915 & 1 & & 0.9573 & 3 & 0.9856 & 5 \\
2 & 0.9829 & 2 & & 0.9659 & 3 & 0.9856 & 5 \\
3 & 0.9801 & 2 & & 0.9687 & 3 & 0.9856 & 5 \\
4 & 0.9801 & 2 & & 0.9687 & 3 & 0.9856 & 5 \\
5 & 0.9829 & 2 & & 0.9687 & 3 & 0.9827 & 5 \\
6 & 0.9858 & 2 & & 0.9687 & 3 & 0.9798 & 5 \\
7 & 0.9886 & 2 & & 0.9687 & 3 & 0.9856 & 5 \\
8 & 0.9829 & 2 & & 0.9687 & 3 & 0.9827 & 5 \\
9 & 0.9801 & 2 & & 0.9602 & 3 & 0.9856 & 4 \\
10 & 0.9829 & 2 & & 0.9659 & 3 & 0.9827 & 5 \\
\end{tabular}
\label{tau.tab}

\medskip

a) Document Ranking task collection

\bigskip

\begin{tabular}{crccrcrc}
 & \multicolumn{2}{c}{All vs. Official} & \hspace{.5in} &
		\multicolumn{4}{c}{Omit Team vs. Official}\\
 & \multicolumn{2}{c}{MAP} & & \multicolumn{2}{c}{MAP} & \multicolumn{2}{c}{Prec(10)} \\
\multicolumn{1}{c}{Trial} & \multicolumn{1}{c}{$\tau$} & \multicolumn{1}{c}{Drop} & &
	\multicolumn{1}{c}{$\tau$} & \multicolumn{1}{c}{Drop} &
	\multicolumn{1}{c}{$\tau$} & \multicolumn{1}{c}{Drop} \\
\cline{2-3} \cline{5-6} \cline{7-8}
1 & 0.9970 & 1 & & 0.9820 & 2 & 0.9939 & 2 \\
2 & 0.9910 & 2 & & 0.9819 & 2 & 0.9939 & 2 \\
3 & 0.9880 & 2 & & 0.9820 & 2 & 0.9939 & 2 \\
4 & 0.9880 & 2 & & 0.9820 & 2 & 0.9939 & 2 \\
5 & 0.9880 & 2 & & 0.9820 & 2 & 0.9939 & 2 \\
6 & 0.9970 & 1 & & 0.9820 & 2 & 0.9939 & 2 \\
7 & 0.9940 & 1 & & 0.9849 & 2 & 0.9939 & 2 \\
8 & 0.9880 & 2 & & 0.9820 & 2 & 0.9939 & 2 \\
9 & 0.9880 & 2 & & 0.9850 & 2 & 0.9939 & 2 \\
10 & 0.9880 & 2 & & 0.9820 & 2 & 0.9939 & 2 \\
\end{tabular}

\medskip

b) Passage Ranking task collection

\end{center}
\end{table*}

The rightmost segment of Table~\ref{tau.tab} gives the $\tau$ and maximum drop values for
the experiments when one participating team is removed from the process.
In these experiments, for each team in turn, we created initial pools consisting
of the MARCO judged documents plus the top-10 documents from all runs except those
runs submitted by the current team.  This pool was fed to the HiCAL process for each of
ten trials where the random number seed for a given trial was the same as in the
all-teams simulation.  As before, we created a trace of the documents that were encountered by
HiCAL, and created a qrels file by taking a prefix of the trace of length equal to
the number of documents judged in the official qrels.  All runs were evaluated
using this trial qrels, and the ranking induced by it was compared to the ranking induced
by the official qrels.
The table reports the smallest $\tau$ and largest maximum drop observed over all teams
for that trial.

\begin{figure*}
\includegraphics[width=3in,height=3in]{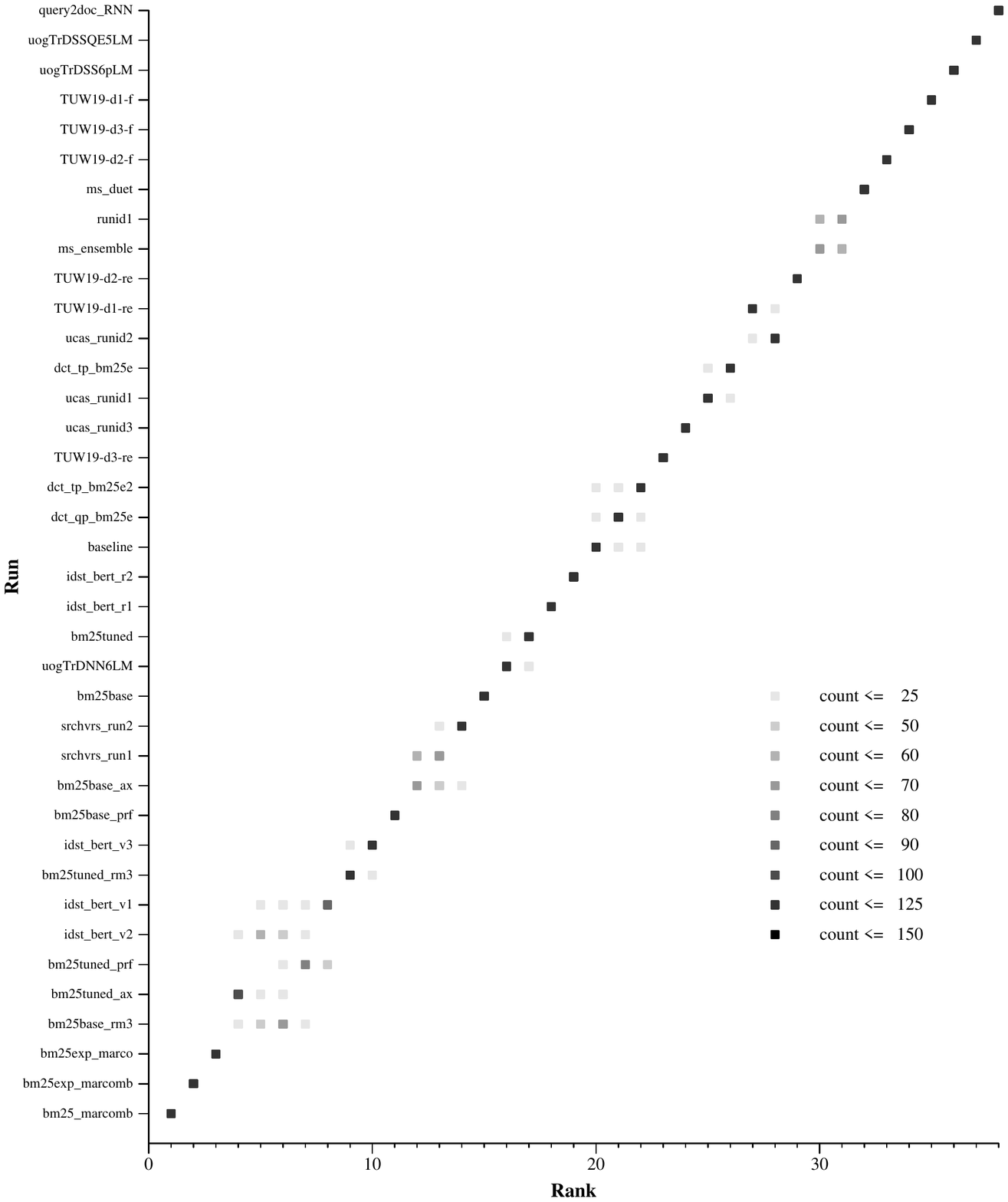}\hfill\includegraphics[width=3in,height=3in]{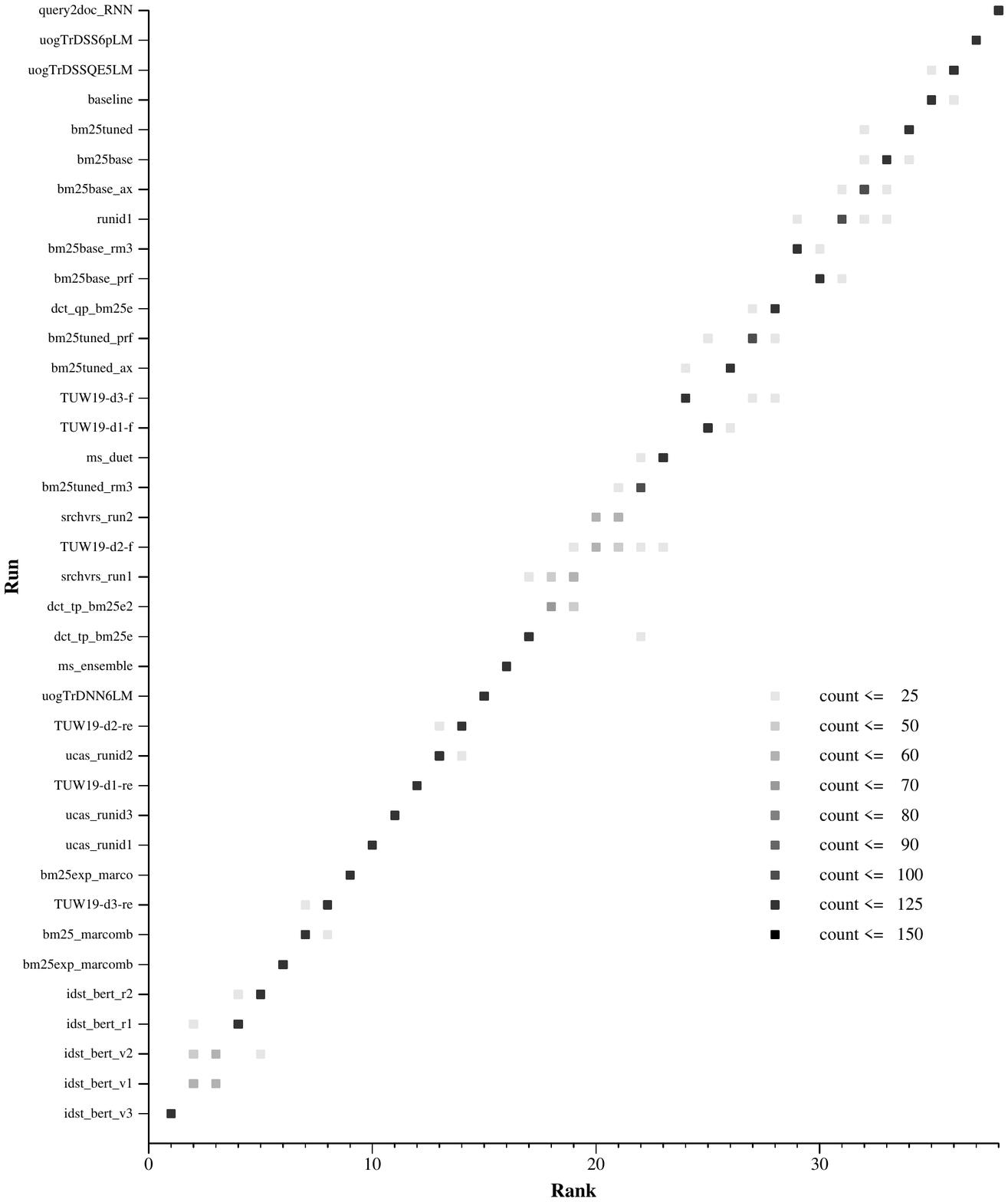}

\vspace*{-.4in}
\hspace*{.75in}Document Ranking collection, MAP \hfill Document Ranking collection, Prec(10) \hspace{.2in}

\includegraphics[width=3in,height=3in]{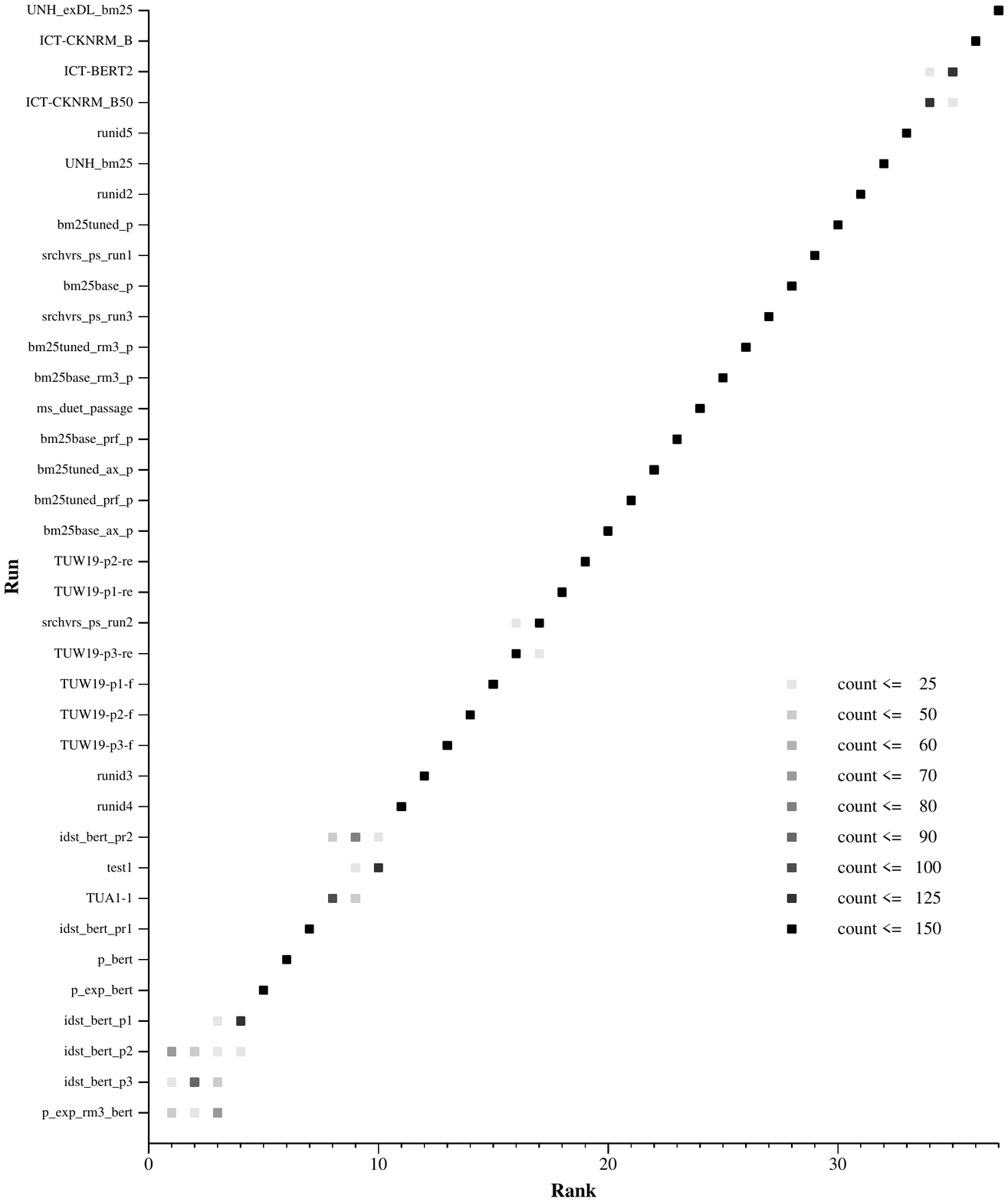}\hfill\includegraphics[width=3in,height=3in]{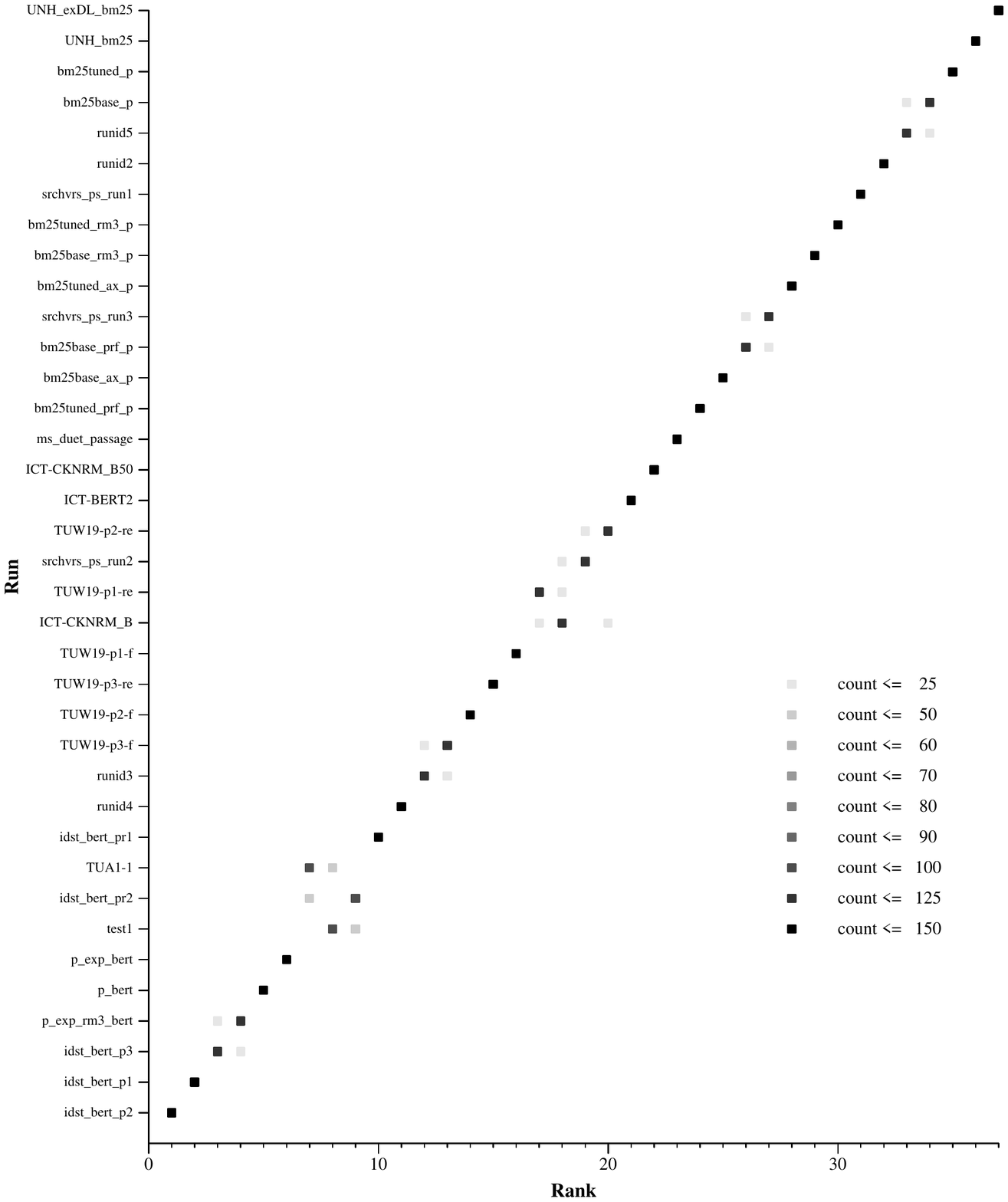}

\vspace*{-.4in}
\hspace*{.75in}Passage Ranking collection, MAP \hfill Passage Ranking collection, Prec(10) \hspace{.2in}
\vspace{.2in}

\caption{Position at which runs ranked over all simulation trials.}
\label{ranks.fig}
\end{figure*}

In general, the ranking of systems is stable, providing support for the contention that
the collections are reusable.  
A more detailed look at the variability in system rankings is given in Figure~\ref{ranks.fig}.
The figure shows a heat map of the number of times a run was ranked
at a given position over all simulation trials (120 trials for the Document Ranking collection
and 130 trials for the Passage Ranking task).
The ranks are plotted on the x-axis and the runs on the y-axis where they are sorted
by their position in the ranking by the official qrels.
The darker a plotted point the more times the run was ranked at that position. 
The figure makes it clear that a large majority of runs have a single dominant rank.
When a run does have change ranks, it moves by a modest amount.

\subsection{Per-topic Budgets}

The qrels created from the simulations for the stability investigation were
constructed to contain exactly the same number of judgments per topic as the
official qrels contains for fair comparisons.
But, of course, no such stopping criterion is available when first building a collection.
The trace of documents encountered by HiCAL in the simulations provides a mechanism
for exploring the effect of different stopping conditions on the final collection.
We construct a qrels by applying a given stopping criterion to a document trace.
For these experiments, all 52 topics start the process and each may be included in the final
qrels if the stopping criterion allows.
Unjudged documents encountered in a simulation are treated as
not relevant.

The simplest stopping criterion is to simply judge an equal number of documents per topic.
Each of the $X$ topics that starts the process gets $\mbox{totalBudget}/X$ judgments, and a topic
is included in the final qrels if at least some minimum number of relevant documents (we use 3)
is found.
The simplicity of this method arises from the fact that topics are independent of one another
once the budget is determined, but equal allotment is known to be sub-optimal for finding
the maximum possible viable topics since ``small'' topics will receive as many judgments
 as ``large'' topics.
An alternative is a strict implementation of the process loosely followed in the track;
we call this the Heuristic stopping criterion.
In the Heuristic simulation experiments here, we capped the number of judgments any topic
can receive at 1000, though that cap was never reached.

Table~\ref{stops.tab} shows the number of judgments required and relative quality of
the qrels created from these different stopping criteria for the two collections 
built in the track.
Note that the only judgments available for these collection is from the Official qrels,
so a method could never find more relevant than in the Official qrels.
The statistics for the Document Ranking task collection are given in the top half of
the table and for the Passage Ranking task collection in the bottom half.
The statistics for the Official qrels is included in the table for reference.
The qrels designated as ``Original Size'' are the same qrels as in 
 the previous experiments above:
pools are built from all runs but ten different trials of the HiCAL process,
corresponding to ten different random number seeds, are tested.
``Budget 400'' and ``Budget 500'' correspond to a constant per-topic budget of 400 and
500 judgments respectively.
\begin{table}
\begin{center}
\caption{\rule[-3mm]{0mm}{8mm} Effect of stopping criteria on qrels quality and number
judgments required.}
\label{stops.tab}
\begin{tabular}{lcccccc}
 & Total & \# Eval & \multicolumn{2}{c}{MAP} & \multicolumn{2}{c}{Prec(10)} \\
Criterion & Judgments & Topics & $\tau$ & Drop & $\tau$ & Drop \\
\hline
Official &  20,157 & 43 & --- & --- & --- & --- \\
Original Size & 20,157 & 43 & 0.9801 & 2 & 1.0000 & 0  \\
Budget 400 & 20,852 & 50 & 0.9316 & 5 & 0.9017 & 8 \\
Budget 500 & 26,052 & 50 & 0.9431 & 3 & 0.9017 & 8 \\
\raisebox{-1.5ex}[0pt]{Heuristic} & \multicolumn{1}{p{1in}}{\centering 17,231.2 \mbox{[17,190--17,262]}} & \raisebox{-1.5ex}[0pt]{38} & \raisebox{-1.5ex}[0pt]{0.9260} & \raisebox{-1.5ex}[0pt]{5} & \raisebox{-1.5ex}[0pt]{0.9565} & \raisebox{-1.5ex}[0pt]{2} \\
\end{tabular}

\medskip

a) Document Ranking task collection

\bigskip

\begin{tabular}{lcccccc}
    & Total & \# Eval & \multicolumn{2}{c}{MAP} & \multicolumn{2}{c}{Prec(10)}\\
Criterion & Judgments & Topics & $\tau$ & Drop & $\tau$ & Drop \\
\hline
Official & 11,904 & 43 & --- & --- & --- & --- \\
Original Size &	11,904 & 43 & 0.9880 & 2 & 1.0000 & 0 \\
Budget 400 & 20,852 & 49 & 0.9880 & 1 & 0.9727 & 3 \\
Budget 500 & 26,052 & 49 & 0.9880 & 1 & 0.9727 & 3 \\
\raisebox{-1.5ex}[0pt]{Heuristic} & \multicolumn{1}{p{1in}}{\centering \mbox{12,721.6} \mbox{[12,712-12,730]}} & \raisebox{-1.5ex}[0pt]{46} & \raisebox{-1.5ex}[0pt]{0.9880} & \raisebox{-1.5ex}[0pt]{1} & \raisebox{-1.5ex}[0pt]{0.9786} & \raisebox{-1.5ex}[0pt]{2} \\
\end{tabular}

\medskip

b) Passage Ranking task collection
\end{center}
\end{table}

The Total Number of Judgments column in the table gives the number of judgments used
over all topics that start the process.
These judgments must be made to determine whether a topic will be included in the final
evaluation set, and so must be accounted for in the budgeting process.
The Number of Evaluation Topics is the number of topics that are included in the final
qrels file based on the criterion's specification.
Original Size qrels always have the same number of judgments as the official qrels
by construction, so the qrels built using that method in each trial has the same
number of topics as the qrels from all other trials, namely the number of topics
in the Official qrels.
Constant budget qrels omit a topic only if the minimum number of relevant documents
for a topic is not found.  While it is possible for qrels created by a constant budget to differ
in the number of topics, for the current collections each trial produced a qrels with
the same number of topics as the other trials.
The Heuristic method omits not only topics with too few relevant documents but topics
with too many relevant as well.  Again, different trials could lead to different
numbers of topics in the qrels, but that did not happen in practice.
The Heuristic method is the only method among those tested that can differ in the number of
documents judged across trials.  For that method, the table reports the mean
number of judgments across the ten trials as well as the minimum and maximum number of
judgments observed in a trial.
The remaining columns in the table give the Kendall's $\tau$ score and maximum drops
for the ranking of systems produced by the test qrels as compared to the ranking produced by
the Official qrels.
As in the experiments above, the value reported is the smallest $\tau$ and largest drop
observed across the ten trials.

The main take-away from the results in Table~\ref{stops.tab} is that the HiCAL process
is very stable across trials and is even robust to differences in stopping conditions
within the ranges tested.
The primary effect of the different stopping conditions is the inclusion or exclusion
of topics affecting mean scores, not differences in individual topic scores.
Averaging effects are the sole explanation for the differences in Prec(10) rankings:
since the top-10 pool was always judged in all conditions, the only difference that can
arise for a Prec(10) ranking is the change in the mean score when a topic is omitted
from the evaluation set.
A large majority of the topics omitted by the Heuristic method were eliminated
by matching the condition $|\mbox{Relevant}| > 0.6|\mbox{Judged}|$ 
once sufficiently many documents were judged (i.e., in step 2 above).

LOU tests and other simulations are dependent on the results submitted to the track, so it
is not possible to say with certainty that a given partially judged collection is reusable.
Nonetheless, the current evidence suggests that the collections built in the Deep Learning track
are high quality ad~hoc collections.

\section{Conclusion}
\label{sec:conclusion}

The TREC 2019 Deep Learning Track introduced two large training datasets, for a document retrieval task and a passage retrieval task, generating two ad hoc test collections with good reusability. For both tasks, in the presence of large training data, this year's non-neural network runs were outperformed by neural network runs. Among the neural approaches, the best-performing runs tended to use transfer learning, employing a pretrained language model such as BERT. In future it will be interesting to confirm and extend these results, understanding what mix of data and multi-stage training lead to the best overall performance. 


We compared reranking approaches to end-to-end retrieval approaches, and in this year's track there was not a huge difference, with some runs performing well in both regimes. This is another result that would be interesting to track in future years, since we would expect that end-to-end retrieval should perform better if it can recall documents that are unavailable in a reranking subtask.

This year there were not many non-neural runs, so it would be important in next year's track to see more runs of all types, to further understand the relative performance of different approaches. Although this year's test collections are of high quality, meaning that they are likely to give meaningful results when reused, overfitting can still be a problem if the test set is used multiple times during the development of a new retrieval approach. The most convincing way to show that a new approach is good is to submit TREC runs. There is no chance of overfitting, or any kind of repeated testing, because the test labels are not generated until after the submission deadline. Through a combination of test collection reuse (from past years) and blind evaluation (submitting runs) the Deep Learning Track is offering a framework for studying ad hoc search in the large data regime. 



\bibliographystyle{plainnat}
\bibliography{bibtex}

\end{document}